\documentclass[aps,prb,reprint,groupedaddress]{revtex4-2}

\usepackage{mathtools}
\usepackage{amssymb}
\usepackage{bm}
\usepackage{graphicx}
\usepackage{booktabs}
\usepackage{array}
\usepackage{hyperref}
\hypersetup{unicode=true,colorlinks=true,linkcolor=blue,citecolor=blue,urlcolor=blue}
\allowdisplaybreaks
\graphicspath{{./Figures/}{./}}

\newcommand{\dd}{\mathrm{d}}
\newcommand{\ee}{\mathrm{e}}
\newcommand{\ii}{\mathrm{i}}
\newcommand{\avg}[1]{\left\langle #1 \right\rangle}

\begin{document}

\title{Quantum Fluctuations and Wigner Dynamics of a Carbon Nanotube Oscillator Driven by an Atomic Force Microscope}

\author{Conrad Silio}
\affiliation{Department of Physics and Astronomy, Tufts University, Medford, MA 02155, USA}

\author{Cristian Staii}
\email{cstaii01@tufts.edu}
\affiliation{Department of Physics and Astronomy, Tufts University, Medford, MA 02155, USA}

\date{\today}

\begin{abstract}
Suspended carbon nanotubes provide a useful platform for studying controlled nanomechanical motion in regimes where environmental fluctuations, memory effects, and quantum coherence can all influence the dynamics. We model the fundamental flexural mode of a carbon nanotube driven by an atomic force microscope as a harmonic quantum Brownian oscillator subject to a prescribed classical force and coupled linearly to a thermal environment. Starting from the generalized Langevin equation, we formulate the reduced dynamics directly in phase space, where the Wigner function evolves through a Gaussian propagator determined by dissipative linear transport, a deterministic force-induced displacement, and bath-induced fluctuations. For a Drude memory kernel, the retarded response is expressed through the three roots of a characteristic cubic equation. In the classical-noise limit, the three independent entries of the finite-time covariance matrix reduce to explicit finite sums over these response poles, eliminating numerical double-time covariance integration. We extend the construction to the full quantum fluctuation spectrum, obtaining a thermal frequency series representation that retains zero-point and low-temperature fluctuations. The asymptotic classical covariance further yields the uncertainty-based consistency criterion $k_BT\geq\hbar\omega_0/2$, which identifies regimes in which the classical equipartition covariance is incompatible with a physical quantum state. The pole construction is also generalized to finite sums of exponential memory kernels, providing an analytic treatment of environments with multiple relaxation times. For an initial two-packet superposition, we examine interference attenuation and the integrated negative part of the Wigner function, which provides a phase space measure of nonclassicality. A deterministic atomic force microscope drive translates the Wigner distribution without changing its integrated negative part, whereas environmental fluctuations suppress nonclassical interference. These results provide an explicit analytical framework for finite-memory quantum Brownian motion of bound oscillators and a tractable phase space description of driven carbon nanotube dynamics relevant to future quantum state preparation and quantum tomography experiments.
\end{abstract}

\maketitle

\section{Introduction}
\label{sec:intro}

Suspended carbon nanotubes (CNTs) combine very small effective masses, high mechanical frequencies, large zero-point displacements, and~experimentally demonstrated high quality factors, making them useful model systems for mesoscopic physics~\cite{Sazonova2004,Huttel2009, Staii2005DNA, Staii2005SGM,
garcia-sanchez_mechanical_2007,tavernarakis_optomechanics_2018, Schneider2014,Samanta2023}.
We recently proposed an atomic force microscope (AFM)-based architecture in which a nearby tip supplies calibrated, time-dependent forces for coherent control and Wigner function tomography of the fundamental CNT flexural mode~\cite{Staii2026}. That work focuses on the anharmonic Duffing/Kerr regime, where spectral anharmonicity can isolate the two~lowest vibrational states. The~present work addresses a complementary question: what can be said analytically about the full phase space evolution when the same mode is treated in the harmonic limit and environmental memory is retained?

The corresponding open system problem is quantum Brownian motion (QBM): a harmonic coordinate linearly coupled to a bath of harmonic oscillators~\cite{CaldeiraLeggett1983,GrabertSchrammIngold,FordLewisOConnell1988,
schlosshauer_decoherence_2007,breuer_theory_2009,Ghosh2024,HomaBernadCsordas2023}.
Hu, Paz, and~Zhang derived an exact reduced master equation for this model under a factorized initial preparation~\cite{HPZ1992}, and~Ford and O'Connell gave a closely related exact solution for the Wigner function based on the quantum Langevin equation~\cite{FordOConnell2001}. A~later analysis by Fleming, Roura, and~Hu recast the problem in a compact phase space form, treated finite-cutoff nonlocal dissipation, and~showed that an external classical force changes the mean while the environmental covariance is determined by the reservoir~\cite{FlemingRouraHu2011}. Importantly, that work also identified an important distinction in converting an integro-differential equation to a time-local master equation: for nonlocal dissipation, the~transition matrix $\Phi(t,\tau)=\Phi(t)\Phi^{-1}(\tau)$ is not equal to the retarded propagator $\Phi(t-\tau)$. Consequently, both the effective force coefficient and the local diffusion coefficients contain additional terms that vanish only in the local limit~\cite{FlemingRouraHu2011}.

More recently, Kondaurov and Polyakov showed that, for~the Caldeira--Leggett model with an external potential that is at most quadratic, the~exact quantum dynamics can be mapped at arbitrary temperature onto a classical non-Markovian stochastic process in phase space~\cite{KondaurovPolyakov2026}. In~their formulation, the~system begins from a correlated thermal equilibrium state, while arbitrary quantum state preparations are incorporated through Wigner-function weights. The~quantum character of the environment is encoded in colored stochastic noise satisfying the full quantum fluctuation--dissipation relation. Their calculations further demonstrate that vacuum and low-temperature fluctuations can produce faster decoherence and rapid short time momentum spreading than classical white noise approximations, emphasizing the importance of retaining the full quantum noise spectrum in low-temperature nanomechanical~systems.

\begin{figure}[t]

\centering
\includegraphics[width=0.95\columnwidth]{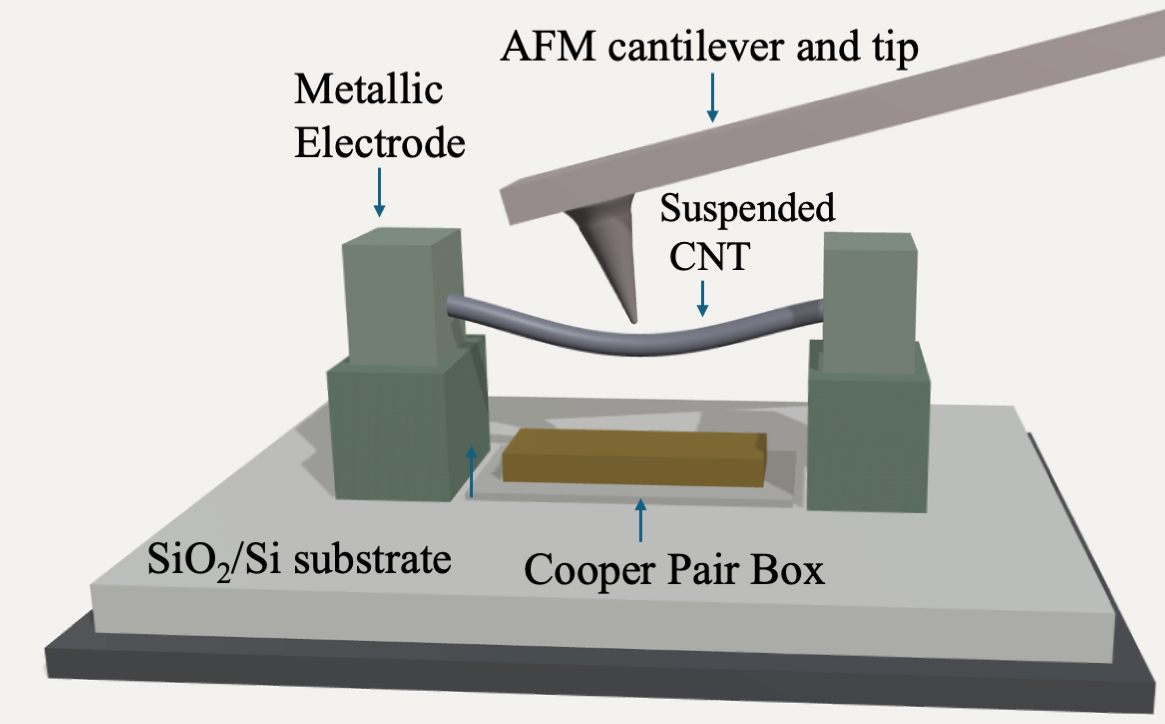}
\caption{Schematic
 illustration of the proposed nanomechanical measurement platform. A~single-walled carbon nanotube (CNT), clamped at both ends and with length $L=100$--$1000$ nm, spans a 200~nm trench and is mechanically actuated by an atomic force microscope (AFM) tip positioned near the CNT. Operation at dilution-refrigerator temperatures ($T\sim 10$ mK) enables coherent control protocols, including Ramsey interferometry and reconstruction of the nanomechanical Wigner function~\cite{Staii2026}. A~Cooper-pair box fabricated on the SiO$_2$/Si substrate is also shown as a representative element for detection of the CNT motion~\cite{blencowe_quantum_2004,Staii2026}.}
\label{fig:setup}
\end{figure}

The present work develops an explicit finite-time phase space solution for the driven \emph{bound} Brownian oscillator with finite environmental memory and applies it to the harmonic CNT mode (Figure~\ref{fig:setup}). Mathematically, the~central contribution is the reduction of the nonlocal bound-oscillator dynamics and its finite-time covariance integrals to explicit pole-based expressions, together with a systematic extension to memory kernels represented by finite sums of exponentials. For~a Drude environment, the~retarded response is expressed through the characteristic poles of the bound oscillator, and~the same structure reduces the finite-time environmental covariance to directly computable analytic expressions. We extend this construction from the classical thermal limit to the full quantum fluctuation spectrum and to finite sums of exponential memory kernels. The~resulting covariance also yields a consistency criterion for determining when the asymptotic classical-noise state is incompatible with the one-mode uncertainty relation. Finally, for~two-packet initial states, we combine the finite-memory attenuation of interference with the integrated Wigner negativity, defined as the integrated magnitude of the negative part of the Wigner function and used here as a phase space measure of nonclassicality. We show that a deterministic AFM force translates the Wigner distribution without changing this integrated negative part. The~framework therefore connects finite-memory dissipation and fluctuations, coherent forcing, and~phase space coherence within a single analytic~description.

These results substantially extend the established exact phase space framework of linear QBM. Ford and O'Connell obtained the Gaussian Wigner propagator for a factorized oscillator--bath preparation and treated Gaussian packets and two-packet superpositions, including the corresponding coherence-attenuation factor~\cite{FordOConnell2001}. The~present problem emphasizes the bound finite-memory oscillator, for~which the restoring force is retained and the response contains both mechanical dynamics and reservoir relaxation. This distinction is important for nanomechanical resonators because the relevant degree of freedom is an oscillatory flexural mode rather than translational Brownian motion. The~externally applied AFM force is incorporated consistently with the known structure of linearly driven QBM, in~which deterministic forcing changes the phase space displacement while the reservoir determines the fluctuation covariance~\cite{FlemingRouraHu2011,QiuQuan2021,breuer_theory_2009}. The~Wigner representation provides a natural setting for this separation and for quantifying nonclassical interference through both fringe visibility and Wigner negativity~\cite{Case2008, bertet_direct_2002, Leibfried1998, karlovets_possibility_2017, WeinbubFerry2018, lee_strong_2023}.

More broadly, these results provide a tractable analytical framework for
quantum Brownian motion of bound oscillators with structured, finite-memory
environments. The~pole representation clarifies how reservoir correlation
times enter both dissipative response and quantum fluctuations, and~the
extension to multiple exponential memory scales makes the method applicable
to a wider class of non-Markovian environments~\cite{Park2024Pseudomode,DAbbruzzo2025, Staii2026Entropy, XuStockburgerAnkerhold2026,
Barandes2025,BarandesKagan2023}. For~nanomechanical resonators, the~framework provides a direct route from experimentally relevant mechanical and reservoir parameters to finite-time phase space dynamics, coherence loss, and~the validity of classical-noise approximations. For~AFM-driven suspended CNTs in particular, it connects calibrated mechanical control with Wigner function evolution and decoherence, providing a quantitative basis for analyzing future experiments on coherent state manipulation and quantum state~tomography.

\section{Driven CNT as a Quantum Brownian~Oscillator}
\label{sec:model}

\subsection{Hamiltonian and Initial~Preparation}

Let $q$ denote the midpoint displacement coordinate of the fundamental CNT
flexural mode, $p$ its conjugate momentum, $m$ its effective mass, and~$\omega_0$ its harmonic frequency. In~the harmonic approximation, the~driven
independent-oscillator Hamiltonian is
~\cite{breuer_theory_2009, schlosshauer_decoherence_2007}

\begin{equation}
\begin{aligned}
H_{\rm tot}(t)
&= \frac{p^2}{2m}+\frac12m\omega_0^2q^2-F(t)q \\
&\quad +\sum_j\left[
\frac{p_j^2}{2m_j}
+\frac12m_j\omega_j^2
\left(q_j-\frac{c_j}{m_j\omega_j^2}q\right)^2
\right].
\end{aligned}
\label{eq:Htot}
\end{equation}
Here,
 $q_j$ and $p_j$ are the coordinate and conjugate momentum of the $j$th
bath oscillator, $m_j$ and $\omega_j$ are its mass and angular frequency,
respectively, and~$c_j$ is the bilinear coupling constant between the CNT
coordinate $q$ and that bath mode. The~shifted coordinate
$q_j-c_jq/(m_j\omega_j^2)$ expresses the fact that, for~a fixed system
displacement $q$, the~equilibrium position of bath oscillator $j$ is displaced
by $c_jq/(m_j\omega_j^2)$. This completed-square form is standard in the
independent-oscillator model of quantum Brownian motion~\cite{CaldeiraLeggett1983,FordLewisOConnell1988,GrabertSchrammIngold}.
Expanding the square separates the bath Hamiltonian, the~bilinear
system--bath interaction, and~the associated counterterm,
\begin{equation}
\begin{aligned}
H_B&=\sum_j\left(
\frac{p_j^2}{2m_j}+\frac12m_j\omega_j^2q_j^2\right),\\
H_I&=-q\sum_j c_jq_j,\\
H_{\rm ct}&=q^2\sum_j\frac{c_j^2}{2m_j\omega_j^2}.
\end{aligned}
\end{equation}
{The} counterterm $H_{\rm ct}$ removes the static frequency renormalization that would otherwise be generated by the bilinear coupling, so that $\omega_0$ specifies the harmonic restoring frequency in the reduced model. This separation is also useful conceptually: $H_I$ mediates the coupling through which the bath produces dissipation and fluctuations, whereas the term $-F(t)q$ represents externally prescribed AFM actuation. Stochastic fluctuations of the AFM tip, if~relevant, would instead be represented by an additional stochastic force or incorporated into the environmental noise model, rather than included in the prescribed deterministic function $F(t)$.

We adopt the factorized preparation used in the HPZ/Ford--O'Connell construction~\cite{HPZ1992, FordOConnell2001},
\begin{equation}
\begin{aligned}
\rho_{\rm tot}(0)
&=\rho_S(0)\otimes\rho_B^{\rm th},\\
\rho_B^{\rm th}
&=\frac{\exp[-H_B/(k_BT)]}
{\mathrm{Tr}_B\{\exp[-H_B/(k_BT)]\}}.
\end{aligned}
\label{eq:factorized}
\end{equation}
{Here,} $\rho_{\rm tot}(0)$ is the density operator of the complete CNT--bath system at the initial time, $\rho_S(0)$ is the initial reduced density operator of the CNT oscillator, and~$\rho_B^{\rm th}$ is the thermal density operator of the uncoupled bath at temperature $T$. Thus, the~bath is initially thermal with respect to the uncoupled bath Hamiltonian and has no initial correlations with $q(0)$ or $p(0)$. Because~this factorized state is not an equilibrium state of the fully coupled Hamiltonian, it gives rise to the familiar initial-slip or initial-jolt behavior~\cite{FordOConnell2001,FlemingRouraHu2011}.

\subsection{Generalized Langevin Equation and Exact~Solution}
\label{sec:langevin}

Eliminating the bath coordinates from the independent-oscillator Hamiltonian
gives the generalized quantum Langevin equation~\cite{FordLewisOConnell1988,GrabertSchrammIngold,HPZ1992},
\begin{equation}
\begin{aligned}
m\ddot q(t)
&+\int_0^t\dd s\,\mu(t-s)\dot q(s)
+m\omega_0^2q(t)\\
&=F(t)+F_B(t)-\mu(t)q(0).
\end{aligned}
\label{eq:GLE}
\end{equation}
where the memory kernel is
\begin{equation}
\mu(t)=
\sum_j\frac{c_j^2}{m_j\omega_j^2}\cos(\omega_j t).
\label{eq:mu}
\end{equation}
The fluctuating bath force is determined by the free evolution of the
uncoupled bath oscillators and can be written as

\begin{equation}
\begin{aligned}
F_B(t)&=\sum_j c_j q_j^{(0)}(t),\\
q_j^{(0)}(t)
&=q_j(0)\cos(\omega_j t)
+\frac{p_j(0)}{m_j\omega_j}\sin(\omega_j t).
\end{aligned}
\label{eq:FB}
\end{equation}
where $q_j^{(0)}(t)$ denotes the evolution of the $j$th bath coordinate
under the uncoupled bath Hamiltonian $H_B$
\cite{FordLewisOConnell1988,GrabertSchrammIngold}. Thus, $F_B(t)$ is an operator determined by the initial bath variables,
whereas $F(t)$ is a prescribed c-number waveform. Both enter the same
linear response equation and are propagated by the same retarded Green
function, but~they play different statistical roles. For~the thermal bath
state in Equation~\eqref{eq:factorized}, $F_B(t)$ has zero mean and its
correlations determine the reservoir-induced fluctuations, whereas $F(t)$
produces a deterministic displacement of the oscillator~\cite{FordLewisOConnell1988,FlemingRouraHu2011}. The~final term
$-\mu(t)q(0)$ in Equation~\eqref{eq:GLE} is the initial-slip contribution
associated with the factorized oscillator--bath \mbox{preparation~\cite{FordOConnell2001,FlemingRouraHu2011}.}

Using the one-sided Fourier--Laplace transform~\cite{FordLewisOConnell1988,GrabertSchrammIngold},
\begin{equation}
\widetilde\mu(s)
=
\int_0^\infty\dd t\,\mu(t)\ee^{\ii st},
\qquad
\operatorname{Im}s>0,
\label{eq:muTransform}
\end{equation}
the symmetrized bath-force correlation function for the thermal bath
state in Equation~\eqref{eq:factorized} is
\begin{equation}
\begin{aligned}
C(t-t')
&\equiv\frac12
\avg{F_B(t)F_B(t')+F_B(t')F_B(t)}\\
&=\frac{1}{\pi}\int_0^\infty\dd\omega\,
\mathrm{Re}\,\widetilde\mu(\omega+\ii0^+)\\
&\quad\times\hbar\omega
\coth\!\left(\frac{\hbar\omega}{2k_BT}\right)
\cos\!\left[\omega(t-t')\right].
\end{aligned}
\label{eq:Ccorr}
\end{equation}
Equation~\eqref{eq:Ccorr} is the quantum fluctuation--dissipation
relation connecting the symmetrized bath fluctuations to the same
memory kernel that determines dissipation~\cite{FordLewisOConnell1988,GrabertSchrammIngold}. {In the present formulation, the~quantum fluctuations of the environment
are encoded in the symmetrized bath-force correlation function}
$C(t-t')${. In~particular, the~factor}
$\hbar\omega\coth[\hbar\omega/(2k_BT)]$
{contains both thermal and zero-point fluctuations. In~the limit}
$T\rightarrow0${, this factor approaches}
$\hbar\omega${, so the bath fluctuations remain nonzero even in the
absence of thermal excitation. These fluctuations enter the reduced Wigner
dynamics through the covariance matrix in Equations~}\eqref{eq:sigmaqqgeneral}--\eqref{eq:sigmappgeneral}{, and~their full finite-temperature form is evaluated explicitly in
Section~\ref{sec:solution}.} The retarded Green function $G(t)$, with~$G(t)=0$ for $t<0$, is defined by
\begin{equation}
\begin{aligned}
m\ddot G(t)
&+\int_0^t\dd u\,\mu(t-u)\dot G(u)
+m\omega_0^2G(t)=0,\\
G(0)&=0,\qquad \dot G(0)=\frac1m.
\end{aligned}
\label{eq:Gdef}
\end{equation}
This Green function is the causal response associated with the
homogeneous generalized Langevin equation~\cite{FordLewisOConnell1988,FordOConnell2001,FlemingRouraHu2011}.
Evaluating Equation~\eqref{eq:Gdef} at $t=0$, the~memory integral vanishes
and $G(0)=0$, giving
\begin{equation}
\ddot G(0)=0.
\label{eq:Gddot0}
\end{equation}
Consequently, for~a regular finite-memory kernel,
\begin{equation}
G(t)=\frac{t}{m}+O(t^3).
\label{eq:Gshort}
\end{equation}
This short time expansion provides a useful consistency check and
determines the leading short-time behavior of the finite-time
covariance derived in Section~\ref{sec:classicalcov}.

The exact initial-value solution can be written in the Heisenberg picture as~\cite{FordOConnell2001,FlemingRouraHu2011}
\begin{align}
q(t)
&=
m\dot G(t)q(0)
+G(t)p(0)
+X(t)
+\eta_q(t),
\label{eq:qsol}\\
p(t)
&=
m^2\ddot G(t)q(0)
+m\dot G(t)p(0)
+m\dot X(t)
+\eta_p(t),
\label{eq:psol}
\end{align}
where
\begin{equation}
\begin{aligned}
X(t)&=\int_0^t\dd u\,G(t-u)F_B(u),\\
\eta_q(t)&=\int_0^t\dd u\,G(t-u)F(u),\\
\eta_p(t)&=m\dot\eta_q(t).
\end{aligned}
\label{eq:XandEta}
\end{equation}
Equations~\eqref{eq:qsol}--\eqref{eq:XandEta} display the linear
structure of the dynamics explicitly. The~motion separates into
propagation of the initial system variables, a~zero-mean fluctuating
contribution generated by the thermal bath, and~a deterministic
displacement generated by the prescribed AFM waveform. The~same
retarded response $G(t)$ propagates both the bath force and the
controlled classical force, while their different statistical
properties give them distinct roles in the reduced dynamics~\cite{FlemingRouraHu2011,QiuQuan2021}. By~introducing the phase space vectors
\begin{equation}
\begin{aligned}
\mathbf z&=(q,p)^{\mathsf T},\\
\boldsymbol{\xi}(t)&=(X(t),m\dot X(t))^{\mathsf T},\\
\boldsymbol{\eta}(t)&=\binom{\eta_q(t)}{\eta_p(t)}.
\end{aligned}
\end{equation}
Equations~\eqref{eq:qsol}--\eqref{eq:psol} become
\begin{equation}
\begin{aligned}
\mathbf z(t)
&=M(t)\mathbf z(0)
+\boldsymbol{\eta}(t)
+\boldsymbol{\xi}(t),\\
M(t)
&=\begin{pmatrix}
m\dot G(t) & G(t)\\
m^2\ddot G(t) & m\dot G(t)
\end{pmatrix}.
\end{aligned}
\label{eq:Mmatrix}
\end{equation}
Thus, the~system phase space dynamics consists of linear transport
generated by $M(t)$ together with additive deterministic and
stochastic displacements. After~averaging over the bath degrees of
freedom, this affine decomposition provides the basis for the reduced
Wigner propagator constructed below~\cite{FordOConnell2001,FlemingRouraHu2011}.

\subsection{Wigner Phase Space~Representation}
\label{sec:wignerintro}

The Wigner function provides a phase space representation of a quantum state
in terms of the same canonical variables, position and momentum, that appear
in classical mechanics~\cite{Case2008,WeinbubFerry2018}.
{The quantum Langevin and Wigner descriptions used here provide
complementary representations of the same underlying system--bath dynamics.
The total Hamiltonian in Equation~}\eqref{eq:Htot} {generates the unitary
evolution of the complete CNT--bath system. In~the Schr\"odinger picture,
the reduced CNT state is}
\begin{equation}
\rho_S(t)=
\operatorname{Tr}_B
\left[
U(t,0)\rho_{\rm tot}(0)U^\dagger(t,0)
\right],
\end{equation}

where $U(t,0)$ {is generated by} $H_{\rm tot}(t)$ \cite{breuer_theory_2009,schlosshauer_decoherence_2007}{. In~the
Heisenberg picture, eliminating the bath coordinates from the equations of
motion generated by the same Hamiltonian gives the generalized quantum
Langevin Equation~}\eqref{eq:GLE}{. The~Wigner function defined below is
the phase space representation of this same reduced density operator}
$\rho_S(t)${. For~the quadratic system--bath Hamiltonian and Gaussian bath
considered here, the~two approaches lead to the same exact reduced dynamics,
represented here by a Gaussian propagator}
\cite{HPZ1992,FordOConnell2001,FlemingRouraHu2011}{.}

{The Wigner representation} is particularly useful for comparing quantum and
classical dynamics and for following the loss of quantum coherence in an
open system. Unlike an ordinary classical probability density, however, the~Wigner function is a quasiprobability distribution. It is real and normalized,
but it need not be positive everywhere. Negative regions and oscillatory
interference structures can encode nonclassical features that cannot, in~general, be represented by an ordinary classical joint probability
distribution~\cite{Case2008,Leibfried1998,WeinbubFerry2018,
bertet_direct_2002,karlovets_possibility_2017,lee_strong_2023}.

For a reduced density operator $\rho_S(t)$, the~Wigner function is defined as
the Fourier transform of the off-diagonal density-matrix elements with respect
to the relative coordinate~\cite{Case2008,WeinbubFerry2018},
\begin{equation}
W(q,p;t)=
\frac{1}{2\pi\hbar}
\int_{-\infty}^{\infty}\dd y\,
\ee^{\ii py/\hbar}
\left\langle q-\frac{y}{2}\right|
\rho_S(t)
\left|q+\frac{y}{2}\right\rangle .
\label{eq:WignerDef}
\end{equation}
With this convention, the~Wigner function is normalized according to~\cite{Case2008,WeinbubFerry2018}
\begin{equation}
\int_{-\infty}^{\infty}\dd q
\int_{-\infty}^{\infty}\dd p\,
W(q,p;t)=1,
\label{eq:Norm}
\end{equation}
and its projections onto the two phase space axes reproduce the usual
position and momentum probability distributions,
\begin{align}
\int_{-\infty}^{\infty}\dd p\,W(q,p;t)
&=\langle q|\rho_S(t)|q\rangle ,
\label{eq:WignerMarginalQ}\\
\int_{-\infty}^{\infty}\dd q\,W(q,p;t)
&=\langle p|\rho_S(t)|p\rangle .
\label{eq:WignerMarginalP}
\end{align}
Thus, the~same phase space function contains both the position and momentum
marginal distributions~\cite{Case2008,WeinbubFerry2018}.

More generally, the~Wigner representation is associated with the Weyl
correspondence between operators and functions on phase space. For~an
operator $\hat A$, its Weyl symbol is defined, using the same Fourier
convention as in Equation~\eqref{eq:WignerDef}, by~\cite{Case2008,WeinbubFerry2018}
\begin{equation}
\widetilde A(q,p)
=
\int_{-\infty}^{\infty}\dd y\,
\ee^{\ii py/\hbar}
\left\langle q-\frac{y}{2}\right|
\hat A
\left|q+\frac{y}{2}\right\rangle .
\label{eq:WeylSymbol}
\end{equation}
Expectation values can then be written as
\begin{equation}
\langle \hat A\rangle
=
\int_{-\infty}^{\infty}\dd q
\int_{-\infty}^{\infty}\dd p\,
W(q,p;t)\,\widetilde A(q,p).
\label{eq:WeylExpectation}
\end{equation}
The appearance of the Weyl symbol is essential: the Wigner function should
not be interpreted as an ordinary joint probability density for
simultaneously measurable position and momentum~\cite{Case2008}.

The phase space formulation is especially natural for open quantum systems.
After the environmental degrees of freedom are traced out, the~system is
generally described by a mixed reduced density operator rather than by a
single wave function. Because~the Wigner function is defined directly from
$\rho_S(t)$, the~same representation applies to both pure and mixed states.
For example, a~coherent superposition of two spatially separated wave packets
produces oscillatory interference fringes in the Wigner distribution.
Environmental coupling can suppress these fine structures and reduce or
eliminate negative regions as coherence is lost~\cite{Leibfried1998,WeinbubFerry2018}. The~Wigner representation therefore
provides a useful description of coherent evolution, environmental
fluctuations, and~the emergence of classical behavior within a common phase
space~framework.

These properties are directly relevant to the CNT oscillator considered
here. The~AFM force contributes a controlled displacement of the mechanical
state, while the environment affects the evolution in two ways: dissipation
modifies the linear transport through $M(t)$, and~fluctuations add Gaussian
broadening through the covariance matrix derived below. The~Wigner function therefore
provides a unified description of controlled motion, dissipative response,
environmental fluctuations, and~the survival or loss of nonclassical
interference. It is also directly relevant to quantum state tomography,
since motional Wigner functions can in principle be reconstructed
experimentally from suitable phase space measurements~\cite{Staii2026, Leibfried1998,WeinbubFerry2018}.

\subsection{Exact Reduced Wigner~Propagator}
\label{sec:solution}

We write the Wigner function as $W(\mathbf z,t)$, with~$\mathbf z=(q,p)^{\mathsf T}$. Because~the initial bath state is Gaussian and
$\boldsymbol{\xi}(t)$ is linear in the initial bath coordinates and momenta,
the bath induces a zero-mean Gaussian contribution to the reduced phase space
dynamics~\cite{FordOConnell2001,FlemingRouraHu2011}. Its symmetrized
covariance matrix is\vspace{-6pt}
\begin{equation}
\begin{aligned}
\Sigma(t)
&=\begin{pmatrix}
\Sigma_{qq}(t) & \Sigma_{qp}(t)\\
\Sigma_{qp}(t) & \Sigma_{pp}(t)
\end{pmatrix},\\
\Sigma_{qq}(t)&=\avg{X^2(t)},\\
\Sigma_{qp}(t)
&=\frac{m}{2}\avg{X(t)\dot X(t)+\dot X(t)X(t)},\\
\Sigma_{pp}(t)&=m^2\avg{\dot X^2(t)}.
\end{aligned}
\label{eq:SigmaDef}
\end{equation}
where the expectation values are taken with respect to the thermal bath state
$\rho_B^{\rm th}$. Thus, $\Sigma_{qq}$ describes the bath-induced position
variance, $\Sigma_{pp}$ the corresponding momentum variance, and~$\Sigma_{qp}$ their symmetrized~covariance.

Following the standard phase space solution of linear quantum Brownian \mbox{motion~\cite{FordOConnell2001,FlemingRouraHu2011},} introduce the characteristic
function using the convention
\begin{equation}
\widetilde W(\mathbf k,t)
=
\int\dd^2z\,
W(\mathbf z,t)
\exp[-\ii\mathbf k^{\mathsf T}\mathbf z],
\qquad
\mathbf k=(k_q,k_p)^{\mathsf T}.
\label{eq:WignerCharacteristic}
\end{equation}
A Gaussian average over the thermal bath then gives~\cite{FlemingRouraHu2011}
\begin{equation}
\begin{aligned}
\widetilde W(\mathbf k,t)
&=\widetilde W\!\left(M^{\mathsf T}(t)\mathbf k,0\right)\\
&\quad\times
\exp\!\left[-\frac12\mathbf k^{\mathsf T}\Sigma(t)\mathbf k\right]\\
&\quad\times
\exp\!\left[-\ii\mathbf k^{\mathsf T}\boldsymbol{\eta}(t)\right].
\end{aligned}
\label{eq:charSolution}
\end{equation}
This is the driven form of the standard linear-QBM phase space solution: the
homogeneous propagator acts on the initial characteristic function, the~environment contributes a Gaussian factor determined by $\Sigma(t)$, and~the prescribed force contributes a phase associated with the deterministic
displacement.

Equivalently, the~reduced Wigner function can be written in terms of a
Gaussian propagator~\cite{FordOConnell2001,FlemingRouraHu2011},
\begin{equation}
W(\mathbf z,t)
=
\int\dd^2z'\,
P(\mathbf z,\mathbf z';t)
W(\mathbf z',0),
\label{eq:propagated}
\end{equation}
where, whenever $\Sigma(t)$ is nonsingular,
\begin{equation}
\begin{aligned}
P(\mathbf z,\mathbf z';t)
&=\frac{1}{2\pi\sqrt{\det\Sigma(t)}}\\
&\quad\times\exp\!\Bigg[-\frac12
\big(\mathbf z-M(t)\mathbf z'-\boldsymbol{\eta}(t)\big)^{\mathsf T}\\
&\qquad\qquad\times\Sigma^{-1}(t)
\big(\mathbf z-M(t)\mathbf z'-\boldsymbol{\eta}(t)\big)
\Bigg].
\end{aligned}
\label{eq:kernel}
\end{equation}
At the initial time, when $\Sigma(0)=0$, Equation~\eqref{eq:kernel} is understood
in the corresponding delta-function limit. The~initial Wigner distribution is therefore transported by $M(t)$,
translated by the deterministic displacement $\boldsymbol{\eta}(t)$, and~broadened by the bath-induced covariance matrix $\Sigma(t)$. The~matrix
$M(t)$ itself also contains the dissipative effect of the environment through
the memory kernel. For~a prescribed noiseless force, $F(t)$ appears only
through $\boldsymbol{\eta}(t)$, whereas $\Sigma(t)$ remains independent of
the driving waveform. This separation between the force-dependent mean and
the reservoir-dependent covariance is a general property of linear QBM with
deterministic forcing~\cite{FlemingRouraHu2011,QiuQuan2021}.

Using the definitions of $X(t)$ and the symmetrized bath-force correlation
function $C(t-t')$, the~bath-induced covariance can be written directly in
terms of the retarded Green function and the noise correlation~\cite{FordLewisOConnell1988,FordOConnell2001,FlemingRouraHu2011}:
\begin{align}
\Sigma_{qq}(t)
&=\int_0^t\dd s\int_0^t\dd s'\,
G(t-s)G(t-s')C(s-s'),
\label{eq:sigmaqqgeneral}\\
\Sigma_{qp}(t)
&=\frac{m}{2}\int_0^t\dd s\int_0^t\dd s'\,
C(s-s')\nonumber\\
&\quad\times\big[G(t-s)\dot G(t-s')
+\dot G(t-s)G(t-s')\big],
\label{eq:sigmaqpgeneral}\\
\Sigma_{pp}(t)
&=m^2\int_0^t\dd s\int_0^t\dd s'\,
\dot G(t-s)\dot G(t-s')C(s-s').
\label{eq:sigmappgeneral}
\end{align}
These expressions are exact within the factorized linear model, provided the
corresponding covariance integrals are well defined for the chosen reservoir
spectrum, and~they provide the starting point for the explicit finite-memory
covariances evaluated in the \mbox{following~sections.}

\subsection{Time-Local Representation and Finite-Memory~Corrections}
\label{sec:localcaution}

Although the nonlocal propagator above is sufficient for the present analysis,
an equivalent time-local Fokker--Planck equation, and~hence a corresponding
time-local master equation, can be constructed on any time interval for which
$M(t)$ is invertible~\cite{HPZ1992,FlemingRouraHu2011}. Introducing the exact
time-local drift matrix
\begin{equation}
A(t)=\dot M(t)M^{-1}(t),
\label{eq:At}
\end{equation}
the first moments satisfy
\begin{equation}
\frac{d}{dt}
\langle \mathbf z(t)\rangle
=
A(t)\langle \mathbf z(t)\rangle
+\mathbf b_{\rm eff}(t),
\label{eq:firstmomentlocal}
\end{equation}
where the effective source vector is
~\cite{FlemingRouraHu2011}
\begin{equation}
\mathbf b_{\rm eff}(t)
=
\dot{\boldsymbol{\eta}}(t)
-A(t)\boldsymbol{\eta}(t).
\label{eq:beff}
\end{equation}
For finite environmental memory, $\mathbf b_{\rm eff}(t)$ is not obtained in
general by simply inserting the original force $F(t)$ into a local
damped-oscillator equation. Rather, it contains the finite-memory correction
to the effective driving term that arises when the original nonlocal dynamics
is rewritten in time-local form~\cite{FlemingRouraHu2011}. More explicitly, using the Fokker--Planck convention~\cite{HPZ1992,FlemingRouraHu2011}
\begin{equation}
\begin{aligned}
\frac{\partial W}{\partial t}
&=-\nabla_{\mathbf z}\!\cdot
\left[\big(A(t)\mathbf z+\mathbf b_{\rm eff}(t)\big)W\right]\\
&\quad+\frac12\sum_{i,j}D_{ij}^{\rm loc}(t)
\frac{\partial^2W}{\partial z_i\partial z_j}.
\end{aligned}
\label{eq:localFP}
\end{equation}

where $D_{ij}^{\rm loc}(t)$ are the matrix elements of the time-dependent
local diffusion matrix $D_{\rm loc}(t)$ in the phase space coordinates
$\mathbf z=(q,p)^{\mathsf T}$.
{Equation~}\eqref{eq:localFP}{ is the exact time-local differential
equation governing the reduced Wigner function on intervals for which}
$M(t)${ is invertible. Equivalently, Equations~}\eqref{eq:propagated}
{and~}\eqref{eq:kernel}{ give its finite-time solution directly in
terms of the response matrix, deterministic displacement, and~bath-induced
covariance. In~the calculations below, we use the finite-time propagator
rather than integrating \mbox{Equation~}\eqref{eq:localFP}}{ as a partial differential
equation, because~it preserves the finite-memory structure explicitly and is
directly applicable to arbitrary initial Wigner functions.} The diagonal elements describe diffusion in
the position and momentum variables, while the off-diagonal elements describe
mixed position--momentum diffusion. The~exact local diffusion matrix associated with the bath-induced covariance
$\Sigma(t)$ is~\cite{FlemingRouraHu2011}
\begin{equation}
D_{\rm loc}(t)
=
\dot\Sigma(t)
-A(t)\Sigma(t)
-\Sigma(t)A^{\mathsf T}(t).
\label{eq:Dloc}
\end{equation}
This relation follows directly from the covariance evolution implied by the
Gaussian propagator.

The time-local representation is therefore completely equivalent to the
nonlocal propagator wherever $M(t)$ is invertible, but~the resulting drift,
driving, and~diffusion coefficients generally retain nontrivial finite-memory
dependence. In~particular, they cannot in general be obtained by replacing
the memory integral in Equation~\eqref{eq:GLE} term by term with an instantaneous
damping term~\cite{FlemingRouraHu2011}. Moreover, if~$\det M(t)$ vanishes at
an isolated time, the~time-local coefficients defined through $M^{-1}(t)$
become singular even though the underlying nonlocal evolution may remain
well defined. For~this reason, we work directly with the nonlocal propagator in
Equation~\eqref{eq:kernel} and the covariance expressions in
Equations~\eqref{eq:sigmaqqgeneral}--\eqref{eq:sigmappgeneral}. This formulation
retains the finite-memory structure explicitly and avoids the intermediate
reconstruction of time-local~coefficients.

\section{Finite-Memory Dynamics for a Drude~Environment}
\label{sec:drude}

\subsection{Response Function and Characteristic~Cubic}

We consider an Ohmic environment with a Lorentz--Drude cutoff. In~the
notation introduced above, the~corresponding exponential memory kernel and
the dissipative part of its one-sided Fourier--Laplace transform are~\cite{FordLewisOConnell1988,GrabertSchrammIngold,FlemingRouraHu2011}
\begin{equation}
\begin{aligned}
\mu(t)&=2m\Gamma_0\omega_c\ee^{-\omega_c t},\qquad t>0,\\
\mathrm{Re}\,\widetilde\mu(\omega+\ii0^+)
&=\frac{2m\Gamma_0\omega_c^2}{\omega_c^2+\omega^2}.
\end{aligned}
\label{eq:drudemu}
\end{equation}
Here, $\Gamma_0$ sets the damping-rate scale, while $\omega_c^{-1}$ is the
reservoir memory time. In~the Markovian, weak-damping limit
$\omega_c\gg\omega_0,\Gamma_0$, the~memory kernel approaches local damping
and the equation of motion reduces to that of a damped harmonic oscillator, with~$Q\simeq\omega_0/(2\Gamma_0)$. For~the CNT application considered here, $\omega_c$ is treated as a
phenomenological environmental parameter rather than as a material constant
fixed by the existing~measurements.

To obtain the retarded response, we introduce the Laplace transform~\cite{FlemingRouraHu2011}
\begin{equation}
\widehat G(\lambda)
=
\int_0^\infty\dd t\,\ee^{-\lambda t}G(t),
\qquad
\operatorname{Re}\lambda>0.
\label{eq:GLaplace}
\end{equation}
Applying this transform to Equation~\eqref{eq:Gdef}, with~$G(0)=0$ and $\dot G(0)=1/m$, gives~\cite{FlemingRouraHu2011}
\begin{equation}
\begin{aligned}
\widehat G(\lambda)
&=\frac{\lambda+\omega_c}{m\,\mathfrak p(\lambda)},\\
\mathfrak p(\lambda)
&=\lambda^3+\omega_c\lambda^2\\
&\quad+\left(\omega_0^2+2\Gamma_0\omega_c\right)\lambda
+\omega_0^2\omega_c.
\end{aligned}
\label{eq:Ghat}
\end{equation}
Thus, the~finite-memory response is governed by the three roots of the
characteristic cubic $\mathfrak p(\lambda)$, as~occurs in the standard
finite-cutoff Ohmic QBM problem~\cite{FlemingRouraHu2011}.

Let $\lambda_1,\lambda_2,\lambda_3$ denote the roots of
$\mathfrak p(\lambda)$. For~distinct roots, inversion of the Laplace
transform by partial fractions gives
\begin{equation}
G(t)
=
\sum_{i=1}^{3}c_i\ee^{\lambda_i t},
\qquad
c_i
=
\frac{\lambda_i+\omega_c}
{m\,\mathfrak p'(\lambda_i)}.
\label{eq:Gpoles}
\end{equation}

For positive $\Gamma_0$, $\omega_0$, and~$\omega_c$, all three roots have
negative real parts, so the retarded response is stable. In~the underdamped
regime, the~cubic contains one real negative root and a complex-conjugate
pair. When the cutoff scale is well separated from the mechanical frequency
and damping scales, the~real root is dominated by reservoir relaxation and
describes a rapidly decaying cutoff-scale transient, while the
complex-conjugate pair describes the damped mechanical oscillation~\cite{FlemingRouraHu2011}. Finite-cutoff response functions with this
additional reservoir-relaxation timescale are well established in exact
treatments of QBM~\cite{FlemingRouraHu2011}.

The pole representation is particularly useful here because the same
exponential structure permits the finite-time covariance integrals to be
reduced to explicit analytic sums. At~isolated parameter values for which two or more roots coincide,
Equation~\eqref{eq:Gpoles} is understood in the continuous repeated-root limit.
In that case, repeated poles generate terms of the form
$t^n e^{\lambda t}$ rather than simple exponentials. For~example, a~double
root produces a contribution $(A+Bt)e^{\lambda t}$.

\subsection{Closed Finite-Time Covariance Matrix in the Classical-Noise~Limit}
\label{sec:classicalcov}

We first consider the classical-noise limit of the fluctuation--dissipation
relation in Equation~\eqref{eq:Ccorr}. When the thermally relevant frequencies
satisfy $\hbar\omega\ll k_BT$, the~quantum thermal factor reduces to~\cite{FordLewisOConnell1988,GrabertSchrammIngold}
\begin{equation}
\hbar\omega
\coth\!\left(\frac{\hbar\omega}{2k_BT}\right)
\longrightarrow
2k_BT.
\label{eq:classicalLimit}
\end{equation}
For the Drude kernel, the~corresponding classical force correlation function
is then~\cite{FordLewisOConnell1988,GrabertSchrammIngold}
\begin{equation}
C(\tau)
=
k_BT\,\mu(|\tau|)
=
K\ee^{-\omega_c|\tau|},
\qquad
K\equiv
2m\Gamma_0\omega_c k_BT.
\label{eq:classicalFDT}
\end{equation}
Because both the retarded response and the force correlation function are
sums of exponentials, the~covariance integrals in
Equations~\eqref{eq:sigmaqqgeneral}--\eqref{eq:sigmappgeneral} can be evaluated
analytically.

Define
\begin{equation}
\mathcal E(a,t)=
\begin{cases}
(\ee^{at}-1)/a, & a\neq0,\\[3pt]
t, & a=0,
\end{cases}
\label{eq:Edef}
\end{equation}
and introduce
\begin{align}
I_{ij}(t)
&\equiv\int_0^t\dd u\int_0^t\dd v\,
\ee^{\lambda_i u+\lambda_j v-\omega_c|u-v|}
\nonumber\\
&=\frac{\mathcal E(\lambda_i+\lambda_j,t)
-\mathcal E(\lambda_i-\omega_c,t)}
{\lambda_j+\omega_c}\nonumber\\
&\quad+\frac{\mathcal E(\lambda_i+\lambda_j,t)
-\mathcal E(\lambda_j-\omega_c,t)}
{\lambda_i+\omega_c}.
\label{eq:Iij}
\end{align}
Here, $\lambda_i$ and $\lambda_j$ are the characteristic roots introduced in
Equation~\eqref{eq:Gpoles}. Although~individual terms associated with the
complex-conjugate poles may be complex, the~complete sums below are real, as~required for the covariance matrix. Substituting the pole representation of $G(t)$ from
Equation~\eqref{eq:Gpoles} into the general covariance expressions yields
\begin{align}
\Sigma_{qq}(t)
&=
K\sum_{i,j=1}^{3}c_i c_j I_{ij}(t),
\label{eq:SigmaqqClosed}\\
\Sigma_{qp}(t)
&=
\frac{mK}{2}
\sum_{i,j=1}^{3}
c_i c_j(\lambda_i+\lambda_j)I_{ij}(t),
\label{eq:SigmaqpClosed}\\
\Sigma_{pp}(t)
&=
m^2K
\sum_{i,j=1}^{3}
c_i c_j\lambda_i\lambda_j I_{ij}(t).
\label{eq:SigmappClosed}
\end{align}
These expressions give the three independent entries of the bath-induced
covariance matrix explicitly at finite time. Together with the response
matrix $M(t)$ and the deterministic displacement
$\boldsymbol{\eta}(t)$ defined above, they make the reduced Wigner
propagator fully explicit for the classical-noise Drude model.
Appendix~\ref{app:covariance} gives the derivation of
\mbox{Equations~\eqref{eq:Iij}--\eqref{eq:SigmappClosed}.} An~independent numerical
validation of the response-pole representation and the closed finite-time
covariance formulas is provided in Appendix~\ref{app:numericalchecks}.

The short-time behavior provides a useful consistency check. Since the
Drude force correlation function is finite at $\tau=0$ and
Equation~\eqref{eq:Gshort} gives $G(t)=t/m+O(t^3)$, direct expansion of the
covariance integrals gives
\begin{equation}
\Sigma_{qq}(t)
=
\frac{K}{4m^2}t^4+O(t^5)
=
\frac{\Gamma_0\omega_c k_BT}{2m}t^4+O(t^5),
\label{eq:t4}
\end{equation}
while
\begin{equation}
\Sigma_{qp}(t)=O(t^3),
\qquad
\Sigma_{pp}(t)=O(t^2).
\label{eq:shortCovOrders}
\end{equation}
{The}
$t^4$
{behavior describes the short time limit of the bath-induced position
variance for the regular finite-memory force correlation and factorized
initial preparation used here. Initial system--environment correlations can
modify the initial-slip transient and the earliest-time covariance evolution}
\cite{RomeroPaz1997,KondaurovPolyakov2026}. {Consequently, the~experimentally relevant short-time behavior can depend
on the preparation of the CNT--environment state and on the reservoir memory
time}
\cite{RomeroPaz1997,KondaurovPolyakov2026}{.}
The $t^3$ position-variance scaling familiar from a white noise description
is recovered only after the finite reservoir correlation time has been
coarse grained. At~fixed finite $\omega_c$, the~strict $t\rightarrow0$
behavior of the bath-induced position variance is instead proportional to
$t^4$. For~a stable bound oscillator coupled to a classical thermal bath satisfying
the fluctuation--dissipation relation, the~response to the initial system
variables decays at long times and the oscillator approaches the classical
thermal covariance~\cite{FordLewisOConnell1988,GrabertSchrammIngold}:
\begin{equation}
\begin{aligned}
\Sigma_{qq}(\infty)&=\frac{k_BT}{m\omega_0^2},\\
\Sigma_{qp}(\infty)&=0,\\
\Sigma_{pp}(\infty)&=mk_BT.
\end{aligned}
\label{eq:equipartition}
\end{equation}

\subsection{Validity of the Classical-Noise~Approximation}
\label{sec:validity}

The replacement leading to Equation~\eqref{eq:classicalFDT} is an additional
high-temperature approximation and is not required for the exact linear QBM
formulation developed in Section~\ref{sec:model}~\cite{FordLewisOConnell1988,GrabertSchrammIngold,Ghosh2024}.
It is justified when $\hbar\omega\ll k_BT$ over the range of frequencies
that contributes appreciably after weighting by both the reservoir spectrum
and the mechanical response. For~a Lorentz--Drude environment, the~dissipative spectrum decreases algebraically rather than terminating at a
sharp cutoff~\cite{GrabertSchrammIngold,FlemingRouraHu2011}. The~condition
$\hbar\omega_c\ll k_BT$ should therefore be interpreted as a useful
scale-separation criterion indicating that the reservoir cutoff scale lies
within the thermal regime, rather than as a uniform classicality condition
extending to arbitrarily large $\omega$. The~mechanical frequency scale must
also be considered. In~particular, $\hbar\omega_0\ll k_BT$ is required when
the oscillator resonance contributes appreciably to the dynamics. By~contrast, the~weaker condition $k_BT\gg\hbar\Gamma_0$ compares the thermal
energy only with the damping scale and is not sufficient to ensure that the
full fluctuation spectrum can be treated classically. More generally, the~treatment of high-frequency fluctuations can depend nontrivially on the
interplay between the high-temperature and large cutoff limits as discussed in Ref.~\cite{Pleasance2026}.

These distinctions can be illustrated using the three representative CNT
parameter sets adapted from Ref.~\cite{Staii2026} and summarized in
Tables
~\ref{tab:physicality} and~\ref{tab:parameters}. At~$T=10$ mK,
\begin{equation}
\frac{k_BT}{h}\simeq208~{\rm MHz}.
\label{eq:thermalFrequency10mK}
\end{equation}
The $5.37$ GHz CNT mode is therefore well outside the classical thermal
regime at its resonance frequency, whereas the $221$ MHz mode lies near the
thermal quantum-to-classical crossover. The~$54$ MHz mode is closer to the
classical regime at its mechanical resonance, although~the quantitative
accuracy of the approximation also depends on the reservoir bandwidth.
Because the existing CNT measurements determine the mechanical frequencies
and quality factors but do not determine a microscopic value of $\omega_c$,
the classical-noise Drude result is best regarded as an analytically
tractable finite-memory reference model. Section~\ref{sec:quantumdrude}
retains the full quantum fluctuation spectrum, while
Section~\ref{sec:physicality} derives a separate uncertainty-based consistency
criterion for determining whether the asymptotic classical covariance is
compatible with a physical quantum state. The~temperature criteria above address the replacement of the quantum
fluctuation spectrum by classical noise, but~they do not remove the
preparation dependence discussed in Section~\ref{sec:model}. In~particular, a~suddenly factorized initial state can retain short time and ultraviolet
sensitivity even when the full quantum bath correlation function is retained~\cite{FordOConnell2001,FlemingRouraHu2011}. Correlated initial preparations
provide a complementary treatment of these effects~\cite{RomeroPaz1997,KondaurovPolyakov2026}. The~full quantum covariance
matrix developed in the following section therefore removes the
classical-noise approximation while retaining the initial-state and
reservoir-spectrum assumptions of the present~model.

\section{Quantum Covariance Matrix for a Drude~Environment}
\label{sec:quantumdrude}

The classical-noise covariance derived in Section~\ref{sec:classicalcov}
replaces the full quantum fluctuation spectrum by its high-temperature
limit and therefore omits zero-point and finite-temperature quantum
corrections. When these contributions are appreciable over the frequencies
sampled by the oscillator and reservoir, the~full thermal factor in the
fluctuation--dissipation relation must be retained~\cite{FordLewisOConnell1988,GrabertSchrammIngold,FlemingRouraHu2011}.
For the Drude environment, the~resulting correlation function can still
be treated analytically by using the standard expansion in bosonic
Matsubara frequencies and combining it with the three response poles
$\lambda_i$ introduced in Equation~\eqref{eq:Gpoles}. Matsubara frequencies
are discrete frequency scales that arise in finite-temperature
quantum statistical mechanics and encode the temperature dependence of
bosonic thermal correlations. Following the standard finite-temperature expansion~\cite{GrabertSchrammIngold}, define
\begin{equation}
\beta\equiv\frac{1}{k_BT},
\qquad
\nu_n=\frac{2\pi n}{\beta\hbar},
\qquad
n=1,2,\ldots ,
\label{eq:matsubara}
\end{equation}
where $\beta$ is the inverse thermal energy and $\nu_n$ are the bosonic
Matsubara frequencies. The~corresponding partial-fraction expansion of
the thermal factor is
~\cite{GrabertSchrammIngold}
\begin{equation}
\hbar\omega
\coth\!\left(\frac{\beta\hbar\omega}{2}\right)
=
\frac{2}{\beta}
+
\frac{4}{\beta}
\sum_{n=1}^{\infty}
\frac{\omega^2}{\omega^2+\nu_n^2}.
\label{eq:cothexpansion}
\end{equation}

Substitution of Equation~\eqref{eq:cothexpansion} and the Drude dissipative
spectrum in Equation~\eqref{eq:drudemu} into the quantum
fluctuation--dissipation relation gives, for~$\tau\neq0$
\cite{FordLewisOConnell1988,GrabertSchrammIngold},
\begin{equation}
\begin{aligned}
C_Q(\tau)
&=K\,\ee^{-\omega_c|\tau|}\\
&\quad+\frac{4m\Gamma_0\omega_c^2}{\beta}
\sum_{n=1}^{\infty}
\frac{\nu_n\ee^{-\nu_n|\tau|}
-\omega_c\ee^{-\omega_c|\tau|}}
{\nu_n^2-\omega_c^2}.
\end{aligned}
\label{eq:CquantumDrude}
\end{equation}
where $C_Q(\tau)$ denotes the full quantum symmetrized bath-force
correlation function and $K=2m\Gamma_0\omega_c k_BT$ is the classical
coefficient defined in Equation~\eqref{eq:classicalFDT}. If~$\nu_n=\omega_c$ for an isolated value of $n$, the~corresponding term
in Equation~\eqref{eq:CquantumDrude} is defined by its continuous~limit.

At equal times, the~quantum force correlator retains the ultraviolet
sensitivity associated with the high-frequency tail of the Drude spectrum.
In particular, $C_Q(0)$ is not finite as an ordinary pointwise force
variance. The~reduced dynamics, however, depends on the force correlation function
through its time convolutions with the retarded response, rather than on
$C_Q(0)$ alone
~\cite{FordOConnell2001,FlemingRouraHu2011}. As~shown below, for~fixed
finite $\omega_c$ and $t>0$, the~Matsubara series obtained after these
time convolutions has summands that decay sufficiently rapidly for the
finite-time covariance to~converge.

To evaluate the time convolutions, generalize the integral introduced in
Equation~\eqref{eq:Iij} by defining, for~$a>0$,
\begin{align}
I_{ij}^{(a)}(t)
&\equiv\int_0^t\dd u\int_0^t\dd v\,
\ee^{\lambda_i u+\lambda_jv-a|u-v|}
\nonumber\\
&=\frac{\mathcal E(\lambda_i+\lambda_j,t)
-\mathcal E(\lambda_i-a,t)}{\lambda_j+a}
\nonumber\\
&\quad+\frac{\mathcal E(\lambda_i+\lambda_j,t)
-\mathcal E(\lambda_j-a,t)}{\lambda_i+a}.
\label{eq:Iija}
\end{align}
Here, $\lambda_i$ and $\lambda_j$ are the same characteristic roots
defined in Equation~\eqref{eq:Gpoles}, and~$\mathcal E(a,t)$ is defined in
Equation~\eqref{eq:Edef}. The~classical Drude integral is recovered as
$I_{ij}(t)=I_{ij}^{(\omega_c)}(t)$. Apparent singularities that may occur
at isolated parameter values in Equation~\eqref{eq:Iija} are understood through
the corresponding continuous~limits.

It is convenient to collect the quantum-noise contribution associated with
each pair of response poles into
\begin{equation}
\begin{aligned}
\mathcal J_{ij}^{Q}(t)
&=K I_{ij}^{(\omega_c)}(t)\\
&\quad+\frac{4m\Gamma_0\omega_c^2}{\beta}
\sum_{n=1}^{\infty}
\frac{\nu_n I_{ij}^{(\nu_n)}(t)
-\omega_c I_{ij}^{(\omega_c)}(t)}
{\nu_n^2-\omega_c^2}.
\end{aligned}
\label{eq:JijQ}
\end{equation}
As in Equation~\eqref{eq:CquantumDrude}, a~term for which
$\nu_n=\omega_c$ is evaluated by taking the corresponding continuous
limit.

Starting from the general Gaussian covariance structure of linear QBM
~\cite{HPZ1992,FordOConnell2001,FlemingRouraHu2011}, and~substituting the
Drude pole representation of $G(t)$ from Equation~\eqref{eq:Gpoles} together
with the quantum-noise integral $\mathcal J_{ij}^{Q}(t)$ defined in
Equation~\eqref{eq:JijQ}, we obtain the three independent entries of the
bath-induced quantum covariance matrix,
\begin{align}
\Sigma_{qq}^{Q}(t)
&=
\sum_{i,j=1}^{3}
c_i c_j\,\mathcal J_{ij}^{Q}(t),
\label{eq:SigmaqqQuantum}\\
\Sigma_{qp}^{Q}(t)
&=
\frac{m}{2}
\sum_{i,j=1}^{3}
c_i c_j(\lambda_i+\lambda_j)\,
\mathcal J_{ij}^{Q}(t),
\label{eq:SigmaqpQuantum}\\
\Sigma_{pp}^{Q}(t)
&=
m^2
\sum_{i,j=1}^{3}
c_i c_j\lambda_i\lambda_j\,
\mathcal J_{ij}^{Q}(t).
\label{eq:SigmappQuantum}
\end{align}

The convergence of this representation follows directly from the
large-$\nu_n$ behavior of Equation~\eqref{eq:Iija}. For~fixed $t>0$,
\begin{equation}
I_{ij}^{(\nu_n)}(t)
=
\frac{2\,\mathcal E(\lambda_i+\lambda_j,t)}{\nu_n}
+
O(\nu_n^{-2}).
\label{eq:IijaAsymptotic}
\end{equation}
Consequently, $\nu_n I_{ij}^{(\nu_n)}(t)$ approaches a finite limit,
whereas the denominator of each term in Equation~\eqref{eq:JijQ} grows as
$\nu_n^2$. The~Matsubara summand therefore decreases as
$O(\nu_n^{-2})$, so the resulting finite-time covariance series converges
for fixed finite $\omega_c$ and $t>0$. This behavior should be
distinguished from the ultraviolet divergence of the unconvolved
equal-time force correlator discussed~above.

In the regime in which $\hbar\omega\ll k_BT$ over the frequencies that
contribute appreciably to the response-weighted noise spectrum, the~Matsubara corrections become negligible and
Equations~\eqref{eq:SigmaqqQuantum}--\eqref{eq:SigmappQuantum} reduce
continuously to the classical-noise expressions of
Section~\ref{sec:classicalcov}. At~lower temperatures, the~additional terms
retain the zero-point and quantum thermal fluctuations absent from the
classical-noise approximation~\cite{FordLewisOConnell1988,GrabertSchrammIngold}.

The representation also separates the two characteristic frequency
structures of the problem. The~poles $\lambda_i$ contain the mechanical
dynamics and finite reservoir memory, whereas the Matsubara frequencies
$\nu_n$ encode the temperature dependence of the quantum noise. The~response matrix $M(t)$ in Equation~\eqref{eq:Mmatrix} and the deterministic
displacement $\boldsymbol{\eta}(t)$ in Equation~\eqref{eq:XandEta} are
unchanged. Replacing the classical correlator by
Equation~\eqref{eq:CquantumDrude} modifies only the bath-induced covariance.
This makes the quantum and classical finite-memory predictions directly
comparable within the same phase space~propagator.

\section{Uncertainty-Based Consistency Criterion for the Classical-Noise~Model}
\label{sec:physicality}

The classical-noise approximation must be assessed in two distinct respects.
It may give quantitatively inaccurate dynamics while still producing a
covariance matrix that is compatible with quantum mechanics, or~it may
predict an asymptotic covariance that violates the uncertainty principle
and therefore cannot represent the covariance of any physical one-mode
quantum state. The~latter possibility can be tested directly from the
classical long-time covariance derived in Section~\ref{sec:classicalcov},
without solving the full quantum~problem.

Let $V_0$ denote the symmetrized covariance matrix of the initial CNT
oscillator state, with~elements
$(V_0)_{ab}=
\frac12\langle
\Delta z_a(0)\Delta z_b(0)+
\Delta z_b(0)\Delta z_a(0)
\rangle$,
where $\Delta\mathbf z=\mathbf z-\langle\mathbf z\rangle$ and
$\mathbf z=(q,p)^{\mathsf T}$. For~the factorized linear QBM dynamics
with deterministic forcing, the~covariance evolves according to
~\cite{FordOConnell2001,FlemingRouraHu2011}
\begin{equation}
V(t)=M(t)V_0M^{\mathsf T}(t)+\Sigma(t).
\label{eq:Vprop}
\end{equation}
The deterministic displacement $\boldsymbol{\eta}(t)$ does not appear in
Equation~\eqref{eq:Vprop} because a deterministic translation changes the first
moments but not the covariance. For~a stable bound oscillator, $M(t)\rightarrow0$ as $t\rightarrow\infty$,
so the asymptotic covariance is determined entirely by the bath-induced
contribution. Using the classical equipartition result in
Equation~\eqref{eq:equipartition}, its determinant is
\begin{equation}
\det V_{\infty}^{\rm cl}
=
\frac{(k_BT)^2}{\omega_0^2}.
\label{eq:detVclassical}
\end{equation}

For a real, symmetric, positive one-mode covariance matrix, compatibility
with the quantum uncertainty relation (in its Robertson--Schr\"odinger form)
is equivalent to~\cite{SimonMukundaDutta1994,breuer_theory_2009}
\begin{equation}
\det V\geq\frac{\hbar^2}{4}.
\label{eq:detphysical}
\end{equation}
Substitution of Equation~\eqref{eq:detVclassical} therefore gives
\begin{equation}
\begin{aligned}
k_BT&\geq\frac{\hbar\omega_0}{2},\\
T&\geq T_{\rm qc}
\equiv\frac{\hbar\omega_0}{2k_B}
=\frac{hf_0}{2k_B}.
\end{aligned}
\label{eq:Tqc}
\end{equation}
where $f_0=\omega_0/(2\pi)$. For~the asymptotic classical equipartition
covariance, this condition is both necessary and sufficient for compatibility
with the one-mode uncertainty relation. It is, however, much weaker than the
condition required for the full fluctuation spectrum to be accurately
described by classical noise. In~particular, satisfying
Equation~\eqref{eq:Tqc} does not imply that
$\hbar\omega\ll k_BT$ throughout the response-weighted reservoir bandwidth.
Moreover, Equation~\eqref{eq:Tqc} is an asymptotic covariance test and does not
by itself establish the physicality of the approximate finite-time map for
all initial~states.

For the three representative CNT frequencies in
Table~\ref{tab:physicality}, the~corresponding thresholds~are shown.

\begin{table*}[t]
\centering
\begin{tabular}{lccccc}
\toprule
\boldmath{$f_0$} &
\boldmath{$T_{\rm qc}$} &
\textbf{Status at} \boldmath{$10$} \textbf{mK} &
\boldmath{$R_{qq}$} &
\boldmath{$R_{qp}$} &
\boldmath{$R_{pp}$}\\
\midrule
5.37 GHz & 129 mK & below threshold & 21.7  & 11.3  & 21.7\\
221 MHz  & 5.30 mK & above threshold & 1.38 & 1.09 & 1.38\\
54 MHz   & 1.30 mK & above threshold & 1.035 & 1.006 & 1.035\\
\bottomrule
\end{tabular}
\caption{{Quantum consistency threshold and representative finite-time
comparison between the full quantum and classical bath-induced covariance
matrices. Here,}
$\Sigma_{ab}^{\rm cl}(t)$
{denotes the classical-noise covariance elements given in Equations~}\eqref{eq:SigmaqqClosed}--\eqref{eq:SigmappClosed}{, while}
$\Sigma_{ab}^{Q}(t)$
{denotes the corresponding full quantum covariance elements in Equations~}\eqref{eq:SigmaqqQuantum}--\eqref{eq:SigmappQuantum}{. The~ratios are defined as}
$R_{ab}(t)=\Sigma_{ab}^{Q}(t)/\Sigma_{ab}^{\rm cl}(t)$
{, and~are evaluated at}
$T=10~\mathrm{mK}${,}
$Q=10^4${,}
$\omega_c/\omega_0=10${, and~}$\omega_0 t=\pi/2${. Because~the reservoir cutoff is not experimentally
known for these CNTs, the~finite-time ratios are illustrative rather than
device-specific predictions.}}
\label{tab:physicality}
\end{table*}

{The} $5.37~\mathrm{GHz}$ {CNT mode at}
$T=10~\mathrm{mK}$  lies well below the threshold in \mbox{Equation~\eqref{eq:Tqc}}, so its asymptotic classical equipartition covariance violates the
one-mode uncertainty bound and cannot represent a physical quantum state.
The
$221~\mathrm{MHz}$
{and}
$54~\mathrm{MHz}$
{modes satisfy this consistency test. The~finite-time comparison in
Table~}\ref{tab:physicality} {shows the same frequency dependence more quantitatively. For~the
representative cutoff used in Table~}\ref{tab:physicality}
{, the~ratios}
$R_{ab}=\Sigma_{ab}^{Q}/\Sigma_{ab}^{\rm cl}$
{quantify the departure of the full quantum covariance from the
classical-noise prediction, with~values near unity indicating close
agreement. The~full quantum covariance differs strongly from its classical-noise counterpart for the}
$5.37~\mathrm{GHz}$
{mode, remains appreciably different for the}
$221~\mathrm{MHz}$
{mode, and~is much closer to the classical result for the}
$54~\mathrm{MHz}$
{mode. In~particular, the~}$221~\mathrm{MHz}$
{mode lies near the regime}
$k_BT\sim\hbar\omega_0${, where quantum corrections to the fluctuation spectrum remain
important. Because~these finite-time ratios depend on the assumed reservoir
cutoff, they provide representative finite-memory comparisons rather than
device-specific predictions. Thus, compatibility with the uncertainty
relation and quantitative validity of the classical-noise approximation
remain distinct requirements, particularly for high-frequency
nanoresonators operated at millikelvin temperatures.}

\section{Extension to Multi-Exponential Memory~Kernels}
\label{sec:multiexp}

To extend the Drude construction to environments with several characteristic
memory times, consider a finite representation of the memory
kernel~\cite{Park2024Pseudomode,XuStockburgerAnkerhold2026},
\begin{equation}
\mu_N(t)
=
\sum_{a=1}^{N}\kappa_a\ee^{-\gamma_a t},
\qquad
\kappa_a>0,
\quad
\gamma_a>0.
\label{eq:multimemory}
\end{equation}
Here, $\gamma_a^{-1}$ is the memory time associated with the $a$th
exponential component, while $\kappa_a$ determines its weight. Equal decay
rates can be combined into a single term, so the $\gamma_a$ may be taken
to be distinct. The~Drude kernel is recovered for $N=1$, with~$\gamma_1=\omega_c$ and $\kappa_1=2m\Gamma_0\omega_c$.

Using the same Laplace transform convention as in the single-Drude
calculation, the~transform of the memory kernel is
\begin{equation}
\widehat\mu_N(\lambda)
=
\sum_{a=1}^{N}
\frac{\kappa_a}{\lambda+\gamma_a}.
\label{eq:muMultiLaplace}
\end{equation}
The generalized Langevin equation then gives the retarded response
~\cite{FlemingRouraHu2011},
\begin{equation}
\begin{aligned}
\widehat G_N(\lambda)
&=\left[m(\lambda^2+\omega_0^2)
+\lambda\sum_{a=1}^{N}
\frac{\kappa_a}{\lambda+\gamma_a}\right]^{-1}\\
&=\frac{\displaystyle\prod_{a=1}^{N}(\lambda+\gamma_a)}
{\mathcal P_{N+2}(\lambda)}.
\end{aligned}
\label{eq:GhatMulti}
\end{equation}
where
\begin{equation}
\begin{aligned}
\mathcal P_{N+2}(\lambda)
&=m(\lambda^2+\omega_0^2)
\prod_{a=1}^{N}(\lambda+\gamma_a)\\
&\quad+\lambda\sum_{a=1}^{N}\kappa_a
\prod_{\substack{b=1\\b\neq a}}^{N}(\lambda+\gamma_b).
\end{aligned}
\label{eq:PolyMulti}
\end{equation}
Thus, an~$N$-exponential memory kernel produces a characteristic polynomial
of degree $N+2$. Let $\lambda_r$, $r=1,\ldots,N+2$, denote the roots of
$\mathcal P_{N+2}(\lambda)$. If~these roots are simple, inversion of the
Laplace transform by partial fractions gives
\begin{equation}
G_N(t)
=
\sum_{r=1}^{N+2}c_r\ee^{\lambda_r t},
\qquad
c_r
=
\frac{\displaystyle\prod_{a=1}^{N}(\lambda_r+\gamma_a)}
{\mathcal P_{N+2}'(\lambda_r)}.
\label{eq:Gmulti}
\end{equation}
As in the single-Drude case, repeated roots are treated by taking the
appropriate continuous limit. A~root of multiplicity greater than one
generates terms proportional to $t^n\ee^{\lambda_r t}$ in addition to
simple~exponentials.

In the classical-noise limit of the fluctuation--dissipation relation,
the force correlation function becomes~\cite{FordLewisOConnell1988,GrabertSchrammIngold}
\begin{equation}
C_N^{\rm cl}(\tau)
=
k_BT\,\mu_N(|\tau|)
=
k_BT
\sum_{a=1}^{N}
\kappa_a\ee^{-\gamma_a|\tau|}.
\label{eq:CmultiClassical}
\end{equation}
Each exponential contribution has the same mathematical form as the
integral evaluated in Equation~\eqref{eq:Iija}. Specifically,
$I_{rs}^{(\gamma_a)}(t)$ denotes the function defined in
Equation~\eqref{eq:Iija}, with~$\lambda_i$ and $\lambda_j$ replaced by
$\lambda_r$ and $\lambda_s$, respectively. Starting from the general Gaussian covariance formulas~\cite{FordOConnell2001,FlemingRouraHu2011}, substitution of the
multi-exponential response and force correlator gives
\begin{align}
\Sigma_{qq}^{(N)}(t)
&=
k_BT
\sum_{a=1}^{N}\kappa_a
\sum_{r,s=1}^{N+2}
c_r c_s I_{rs}^{(\gamma_a)}(t),
\label{eq:SigmaqqMulti}\\
\Sigma_{qp}^{(N)}(t)
&=
\frac{m k_BT}{2}
\sum_{a=1}^{N}\kappa_a
\sum_{r,s=1}^{N+2}
c_r c_s(\lambda_r+\lambda_s)
I_{rs}^{(\gamma_a)}(t),
\label{eq:SigmaqpMulti}\\
\Sigma_{pp}^{(N)}(t)
&=
m^2 k_BT
\sum_{a=1}^{N}\kappa_a
\sum_{r,s=1}^{N+2}
c_r c_s\lambda_r\lambda_s
I_{rs}^{(\gamma_a)}(t).
\label{eq:SigmappMulti}
\end{align}
These expressions reduce exactly to the single-Drude results of
Section~\ref{sec:classicalcov} when $N=1$. The~response to a prescribed AFM
waveform is obtained by replacing $G(t)$ with $G_N(t)$ in the convolution
defining the deterministic displacement, with~the corresponding response
matrix constructed from $G_N(t)$ in the same way as $M(t)$ in
Equation~\eqref{eq:Mmatrix}.

The resulting structure shows why the covariance remains analytically
tractable for this class of environments. A~finite sum of exponential
memory terms produces a rational Laplace-domain response with a finite set
of characteristic poles, while each exponential contribution to the
classical force correlation function is reduced to the previously defined
function $I_{rs}^{(a)}(t)$. The~method therefore accommodates several
distinct reservoir relaxation times without discretizing the bath or
reintroducing numerical double-time covariance integrals~\cite{Park2024Pseudomode,XuStockburgerAnkerhold2026}.

The same construction extends to the full quantum fluctuation spectrum.
Each exponential component of the memory kernel contributes its reservoir
decay scale $\gamma_a$, while the thermal factor introduces the bosonic
Matsubara frequencies $\nu_n$ discussed in Section~\ref{sec:quantumdrude}.
The resulting quantum covariance matrix is therefore organized in terms of the
finite set of response poles $\lambda_r$ and reservoir decay rates
$\gamma_a$, together with the Matsubara frequency series. This preserves
the analytic structure of the single-Drude calculation while allowing for
environments with multiple characteristic memory times to be treated
within the same phase space framework~\cite{GrabertSchrammIngold,Park2024Pseudomode,
XuStockburgerAnkerhold2026}.

\section{AFM Actuation and Representative CNT~Parameters}
\label{sec:afmdrive}

\subsection{Response to a Continuous-Wave~Force}

To connect the finite-memory response with the AFM actuation proposed in
Ref.~\cite{Staii2026}, consider a sinusoidal force switched on at $t=0$,
as in the continuous-wave driving configuration discussed there~\cite{Staii2026}
\begin{equation}
F(t)=F_0\cos(\omega_d t)\,\Theta(t),
\label{eq:cwdrive}
\end{equation}
where $F_0$ is the force amplitude, $\omega_d$ is the drive frequency,
and $\Theta(t)$ is the Heaviside step function. {Equation~}\eqref{eq:cwdrive}
{treats the AFM actuation as a prescribed noiseless force. In~an experiment,
thermal motion of the AFM cantilever, voltage fluctuations, or~tip-position
fluctuations can generate an additional stochastic force on the CNT} \cite{Staii2026}{. If~this
contribution is statistically independent of the thermal reservoir, its force
correlation function adds to}
$C(t-t')$, and~its contribution to the covariance is obtained directly from
\mbox{Equations~\eqref{eq:sigmaqqgeneral}--\eqref{eq:sigmappgeneral}}{. Such AFM-induced noise would therefore produce additional phase space
broadening and could reduce nonclassical interference, in~contrast to the
prescribed deterministic force considered here, which produces only a phase
space displacement.} Using the exact deterministic force convolution of the linear QBM
solution~\cite{FlemingRouraHu2011} and substituting the Drude pole
expansion in Equation~\eqref{eq:Gpoles} gives
\begin{equation}
\begin{aligned}
\eta_q(t)
&=F_0\sum_{i=1}^{3}c_i
\frac{\lambda_i\ee^{\lambda_i t}
-\lambda_i\cos(\omega_d t)}{\lambda_i^2+\omega_d^2}\\
&\quad+F_0\sum_{i=1}^{3}c_i
\frac{\omega_d\sin(\omega_d t)}{\lambda_i^2+\omega_d^2},\\
\eta_p(t)&=m\dot\eta_q(t).
\end{aligned}
\label{eq:etaqclosed}
\end{equation}
Here, $\lambda_i$ and $c_i$ are the same Drude response poles and
residues introduced in Equation~\eqref{eq:Gpoles}. For~a stable response,
the terms proportional to $\ee^{\lambda_i t}$ decay and describe the
transient part of the ring-up, while the sinusoidal terms determine
the long-time driven motion. Although~individual pole contributions
can be complex in the underdamped regime, the~complex-conjugate pair
combines to give a real~displacement.

In the Markovian weak-damping limit,
$\omega_c\gg\omega_0,\Gamma_0$, and~for near-resonant driving
$\omega_d\simeq\omega_0$, the~steady-state displacement amplitude
reduces to the standard damped-oscillator result
~\cite{Staii2026}
\begin{equation}
A_{\rm ss}
\simeq
\frac{F_0}{2m\Gamma_0\omega_0}
=
Q\frac{F_0}{m\omega_0^2},
\label{eq:Ass}
\end{equation}
where $Q\simeq\omega_0/(2\Gamma_0)$ in the damping convention used
here. Equation~\eqref{eq:etaqclosed}, by~contrast, retains the finite
reservoir correlation time and therefore gives both the transient and
long-time AFM response without taking the Markovian~limit.

\subsection{Representative CNT Parameter~Sets}
\label{sec:parameters}

Table~\ref{tab:parameters} collects representative parameter sets adapted
from the CNT tomography proposal in Ref.~\cite{Staii2026}. They are model
estimates based on a doubly clamped Euler--Bernoulli description and assumed
operating conditions rather than measurements of a particular device. As~used in Ref.~\cite{Staii2026}, the~zero-point displacement is
$x_{\rm zpf}=\sqrt{\hbar/(2m\omega_0)}$, the~thermal occupation is
$\bar n_{\rm th}=[\exp(\hbar\omega_0/k_BT)-1]^{-1}$, and, with~the damping convention used here, the~mechanical energy relaxation time is $T_1=Q/\omega_0$. The~quantity $F_{\pi/2}$ is the $\pi/2$-pulse
force estimate introduced in Ref.~\cite{Staii2026} for a
$0.1\,\mu{\rm s}$ control pulse in the anharmonic two-level operating
regime. It is included here as a representative scale for the AFM force
that can be applied to the~CNT.

\begin{table*}[t]

\centering
\begin{tabular}{lcccccc}
\toprule
\boldmath{$L$} \textbf{(nm)} & \boldmath{$f_0$} & \boldmath{$x_{\rm zpf}$} \textbf{(pm) }& \boldmath{$m$} \textbf{(ag)} &
\boldmath{$F_{\pi/2}$} \textbf{(fN)} & \boldmath{$T_1$} \textbf{(}\boldmath{$\mu$}\textbf{s)} & \boldmath{$\bar n_{\rm th}$}\\
\midrule
100  & 5.37 GHz & 2.14 & 0.341 & 0.77 & 0.30 & $6\times10^{-12}$\\
500  & 221 MHz  & 4.79 & 1.66  & 0.35 & 7.4  & 0.53\\
1000 & 54 MHz   & 6.78 & 3.38  & 0.24 & 29.5 & 3.4\\
\bottomrule
\end{tabular}
\caption{Representative CNT parameter sets adapted from
Ref.~\cite{Staii2026}. The~estimates use $Q=10^4$ and $T=10$ mK.
$F_{\pi/2}$ is the force scale for a $0.1\,\mu$s $\pi/2$ control pulse
in the anharmonic two-level model of Ref.~\cite{Staii2026}, and~$\bar n_{\rm th}$ is the equilibrium thermal occupation before active
cooling. The~values are model estimates rather than measured properties
of a single device.}

\label{tab:parameters}
\end{table*}

The reservoir cutoff $\omega_c$ is not included in
Table~\ref{tab:parameters} because a microscopic CNT-specific memory time
is not presently established by the available mechanical parameters.
Consequently, values of $\omega_c/\omega_0$ used in numerical examples
should be interpreted as representative finite-memory regimes rather than
experimentally fitted material parameters. This separation allows for the CNT
mechanical parameters to be specified independently while the influence of
reservoir memory is explored as a phenomenological environmental~effect.

\section{Wigner Dynamics, Coherence Loss, and~the No-Drive~Limit}
\label{sec:cat}

The phase space propagator is particularly informative for an initial state
that contains interference structure. Following the standard two-packet
construction used in quantum Brownian motion~\cite{FordOConnell2001}, we consider a coherent superposition of two
minimum-uncertainty Gaussian packets separated by a distance $d$,
\begin{equation}
\begin{aligned}
\psi(q,0)
&=\frac{1}{(8\pi\sigma^2)^{1/4}\sqrt{1+S}}\\
&\quad\times\Bigg[
\exp\!\left(-\frac{(q-d/2)^2}{4\sigma^2}\right)\\
&\qquad+\exp\!\left(-\frac{(q+d/2)^2}{4\sigma^2}\right)
\Bigg].
\end{aligned}
\label{eq:catpsi}
\end{equation}
where $\sigma$ is the position standard deviation of each packet and
\begin{equation}
S=\exp\!\left[-\frac{d^2}{8\sigma^2}\right]
\end{equation}
is their overlap. Using the definition of the Wigner function in Equation~\eqref{eq:WignerDef}, the~corresponding initial Wigner distribution is
~\cite{FordOConnell2001}
\begin{equation}
\begin{aligned}
W(q,p;0)
&=\frac{1}{2(1+S)}\Bigg[
W_+(q,p)+W_-(q,p)\\
&\quad+\frac{2}{\pi\hbar}
\exp\!\left(-\frac{q^2}{2\sigma^2}
-\frac{2\sigma^2p^2}{\hbar^2}\right)
\cos\!\left(\frac{pd}{\hbar}\right)
\Bigg].
\end{aligned}
\label{eq:catW0}
\end{equation}
where
\begin{equation}
W_\pm(q,p)=
\frac{1}{\pi\hbar}
\exp\!\left[
-\frac{(q\mp d/2)^2}{2\sigma^2}
-\frac{2\sigma^2p^2}{\hbar^2}
\right].
\end{equation}
The two positive lobes correspond to the individual packets, while the
oscillatory term gives the phase space interference pattern. Applying the Gaussian propagator of Section~\ref{sec:solution}, the~phase space
centers of the two packets evolve as~\cite{FlemingRouraHu2011}
\begin{equation}
\bar{\mathbf z}_{\pm}(t)
=
M(t)
\binom{\pm d/2}{0}
+
\boldsymbol{\eta}(t),
\label{eq:catcenters}
\end{equation}
so both branches are transported by the same linear response and translated
by the same deterministic displacement. Following the standard two-packet coherence analysis~\cite{FordOConnell2001}, a~convenient diagnostic of coherence loss is
obtained from the position marginal. For~the minimum-uncertainty packets
in Equation~\eqref{eq:catpsi}, the~position variance of either propagated
packet is~\cite{FordOConnell2001}:
\begin{equation}
\begin{aligned}
\langle\Delta q^2(t)\rangle
&=m^2\dot G^2(t)\sigma^2
+\frac{\hbar^2G^2(t)}{4\sigma^2}\\
&\quad+\Sigma_{qq}(t).
\end{aligned}
\label{eq:packetvariance}
\end{equation}
The corresponding attenuation factor of the interference term is (see Appendix~\ref{app:cat}):
\begin{equation}
a(t)=
\exp\!\left[
-\frac{d^2\Sigma_{qq}(t)}
{8\sigma^2\langle\Delta q^2(t)\rangle}
\right].
\label{eq:attenuation}
\end{equation}
Equation~\eqref{eq:attenuation} has the same functional form as the
two-packet attenuation factor derived by Ford and O'Connell~\cite{FordOConnell2001}. Here, however, the~bath-induced variance
$\Sigma_{qq}(t)$ is evaluated using the finite-memory bound-oscillator
response derived above. In~the classical Drude model,
Equation~\eqref{eq:SigmaqqClosed}, together with the pole representation in
Equation~\eqref{eq:Gpoles}, makes Equation~\eqref{eq:attenuation} fully explicit.
For the quantum Drude model, the~same expression applies with
$\Sigma_{qq}(t)$ replaced by $\Sigma_{qq}^{Q}(t)$ from
Equation~\eqref{eq:SigmaqqQuantum}. The~free-particle Ohmic limiting case of the response, covariance, and~two-packet attenuation factor is verified numerically in
Appendix~\ref{app:numericalchecks}.

Figure~\ref{fig:finite_memory} illustrates how the reservoir correlation
time modifies both interference attenuation and bath-induced spreading.
Panel (a) shows the attenuation coefficient $a(t)$ for two representative
finite-memory regimes together with the Markovian white noise limit, while
panel (b) shows the corresponding normalized position covariance
$m\omega_0^2\Sigma_{qq}(t)/(k_BT)$. As~$\omega_c$ increases, the~reservoir
correlation time $\omega_c^{-1}$ becomes shorter and the finite-memory
curves approach the Markovian result. For~smaller $\omega_c$, memory effects
modify both the accumulation of bath-induced position fluctuations and the
attenuation of the two-packet~interference.

\begin{figure*}[t]

\includegraphics[width=0.95\textwidth]{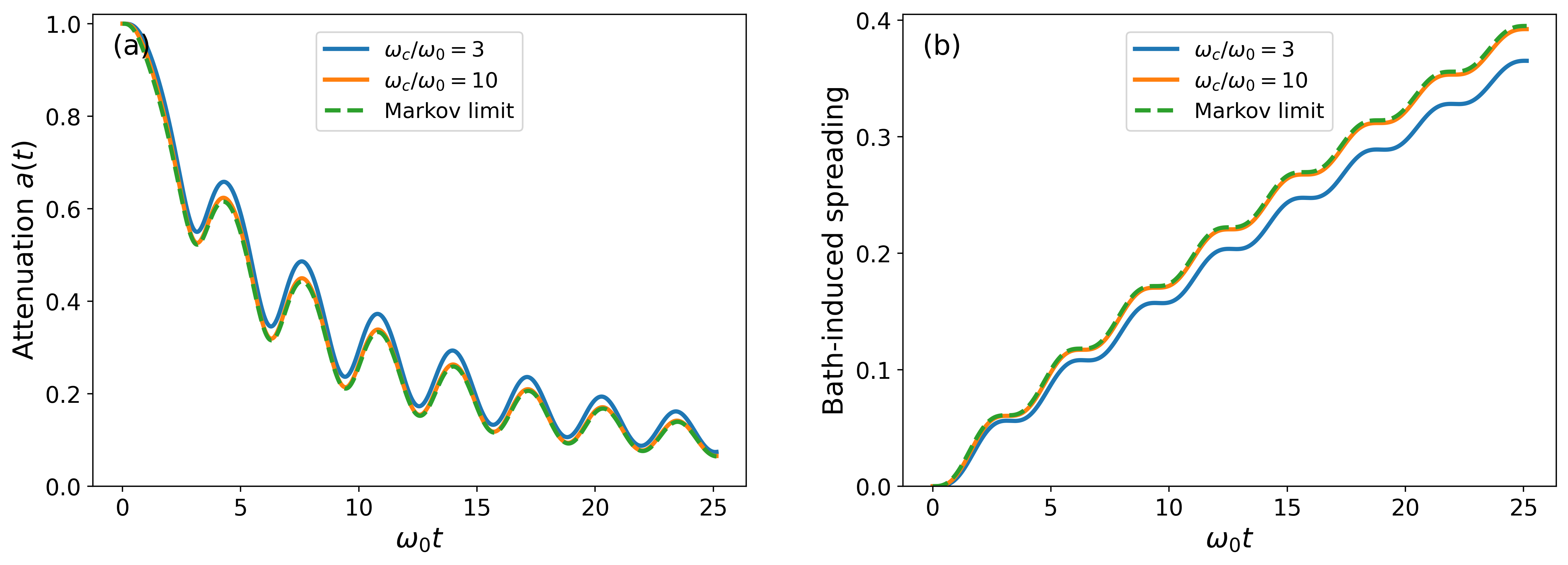}
\caption{Finite-memory corrections to Brownian spreading and coherence
attenuation. The~Drude bath is shown for cutoff ratios
$\omega_c/\omega_0=3$ and $10$, together with the Markov white noise
limit (dashed). The~calculations use the illustrative dimensionless
parameters $m=\omega_0=\hbar=k_BT=1$ and $Q=50$. These values are chosen
to display finite-memory effects and are not intended to represent the
high-temperature regime discussed in Section~\ref{sec:validity}. For~the attenuation coefficient, the~two-packet parameters are $d=3.5$
and $\sigma=0.6$, as~in the Wigner function illustration in
Figures~\ref{fig:drivencat} and~\ref{fig:nodrive}. Panels (\textbf{a} and \textbf{b}) show the attenuation coefficient $a(t)$ and the corresponding normalized
bath-induced position covariance
$m\omega_0^2\Sigma_{qq}(t)/(k_BT)$, respectively. As~the cutoff increases,
the finite-memory curves approach the Markovian result.}
\label{fig:finite_memory}
\end{figure*}

A complementary phase space measure of Wigner nonclassicality is the
integrated Wigner negativity~\cite{KenfackZyczkowski2004},
\begin{equation}
\mathcal N_W(t)=
\frac12
\left[
\int\dd q\,\dd p\,|W(q,p;t)|-1
\right].
\label{eq:Wneg}
\end{equation}
Using the normalization in Equation~\eqref{eq:Norm}, this quantity is equal to
the integrated magnitude of the negative part of the Wigner function,
\begin{equation}
\mathcal N_W(t)
=
\int_{W<0}\dd q\,\dd p\,|W(q,p;t)|.
\end{equation}
Its behavior under deterministic displacement and environmental Gaussian
propagation is analyzed in Section~\ref{sec:negativity}.

\begin{figure*}[t]
\centering
\includegraphics[width=0.95\textwidth]{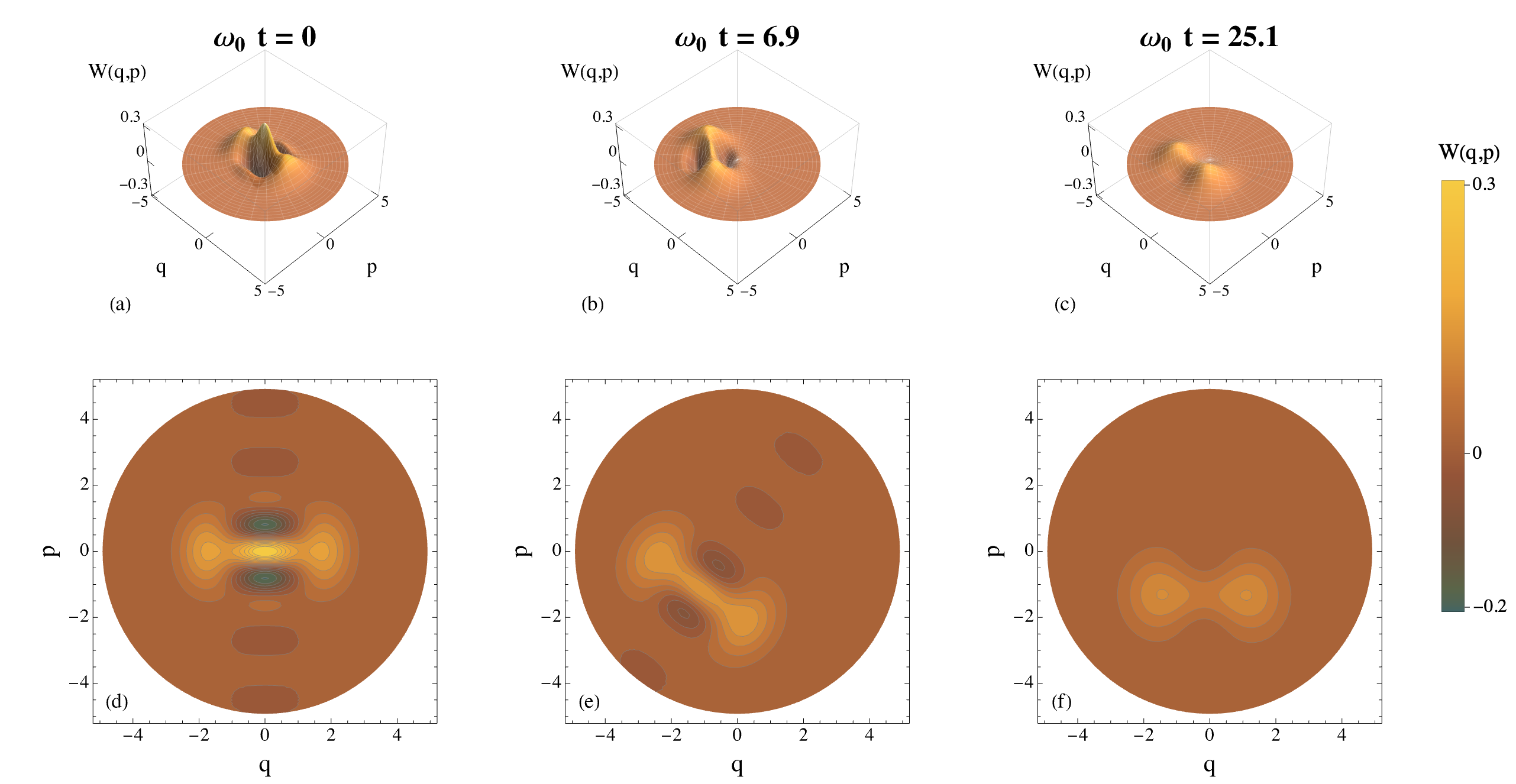}
\caption{Illustration
 of the two-packet Wigner evolution in the presence of
both the finite-memory bath and a prescribed drive. The~top row shows
$W(q,p)$ as a three-dimensional surface and the bottom row shows the
corresponding contour plots at $\omega_0t=0$, $6.9$, and~$25.1$. The~initially negative interference structure is progressively smoothed by the
environmental convolution while the deterministic force displaces the
distribution in phase space. The~calculation is shown in natural oscillator
units $m=\omega_0=\hbar=1${, with~}$Q=50${,}
$\omega_c/\omega_0=10${,}
$k_BT/(\hbar\omega_0)=1${,}
$d=3.5${, and~}$\sigma=0.6${. The~prescribed AFM force is a Gaussian pulse with}
$F_0=1.5${,}
$t_0=3.0${, and~}$\tau_p=0.5${. The~choice}
$Q=50$
{provides enhanced dissipation so that the loss of fine structure is
visible on a few-period timescale.}}
\label{fig:drivencat}
\end{figure*}

Figure~\ref{fig:drivencat} illustrates the combined effects of the prescribed
AFM drive and the finite-memory environment on the Wigner distribution. The~deterministic drive translates the two-packet state through phase space,
while the reservoir-induced evolution modifies the distribution through
dissipative transport and fluctuation-induced broadening. At~the displayed
times, the~negative regions associated with the initial interference pattern
become increasingly smoothed, corresponding to a reduction of the integrated
Wigner negativity relative to its initial value. As~shown more generally in
Section~\ref{sec:negativity}, the~deterministic displacement itself does not
change $\mathcal N_W(t)$.

\begin{figure*}[t]
\centering
\includegraphics[width=0.95\textwidth]{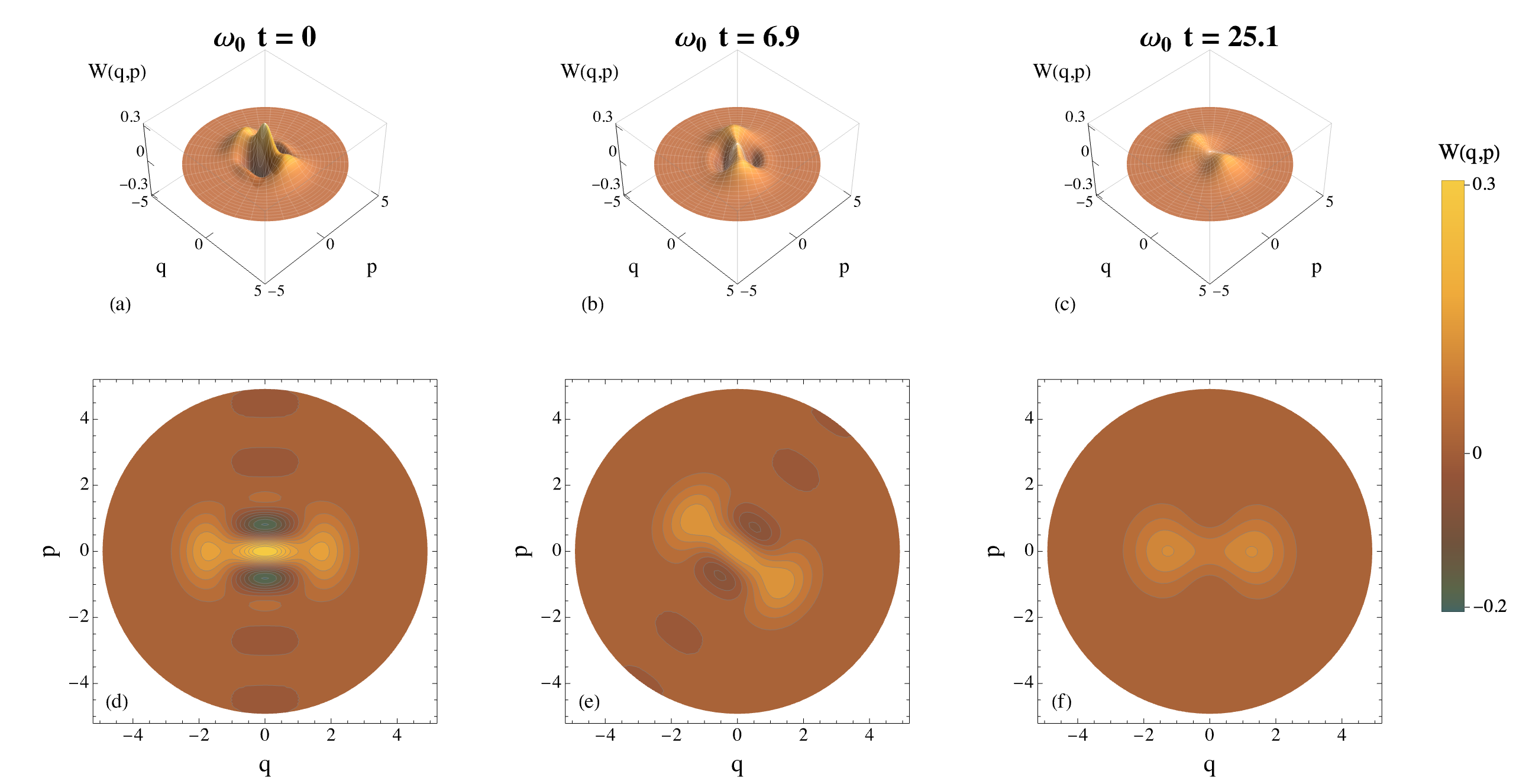}
\caption{Illustrative
 example of the no-drive evolution of the two-packet
state. The~bath and initial state are of the same type as in
Figure~\ref{fig:drivencat}, but~$F(t)=0$ and therefore
$\boldsymbol{\eta}(t)=0$. {The calculation uses}
$m=\omega_0=\hbar=1${,}
$Q=50${,}
$\omega_c/\omega_0=10${,}
$k_BT/(\hbar\omega_0)=1${,}
$d=3.5${, and~}$\sigma=0.6${.}
The negative interference structure is suppressed by the bath without the
additional drive-induced displacement of the phase space distribution. This
is the limiting case associated with the standard no-drive QBM~propagator.}
\label{fig:nodrive}
\end{figure*}

Figure~\ref{fig:nodrive} shows the corresponding no-drive evolution for the
same type of initial two-packet state and bath, thereby isolating the effect
of the environment in the absence of a deterministic phase space
displacement. The~no-drive limit also provides a useful consistency check and
connects the present formulation to the standard factorized-state QBM
propagator. Setting $F(t)=0$ gives $\boldsymbol{\eta}(t)=0$, so the propagator
of Section~\ref{sec:solution} reduces to the corresponding Gaussian no-drive
form~\cite{FordOConnell2001}. The~reservoir continues to produce dissipative
transport and fluctuation-induced broadening, but~there is no externally
generated translation of the phase space distribution. The~two-packet
evolution in this limit therefore isolates the environmental effects on
spreading, interference attenuation, and~Wigner~nonclassicality.

{The Wigner distributions in Figures~}\ref{fig:drivencat} {and~}\ref{fig:nodrive}  are obtained by direct evaluation of the finite-time propagation formula in \mbox{Equation~\eqref{eq:propagated}}{, using the Gaussian kernel in Equation~}\eqref{eq:kernel} {and the initial two-packet Wigner function in Equation~}\eqref{eq:catW0}{. This propagation is equivalent to solving the time-local Fokker--Planck equation for the Wigner function,
Equation~}\eqref{eq:localFP}, {wherever} $M(t)$ {is invertible.} Comparing Figures~\ref{fig:drivencat} and~\ref{fig:nodrive} makes the separation
between deterministic control and reservoir-induced evolution particularly
transparent. The~prescribed AFM force changes the phase space location of the
Wigner distribution but, by~itself, does not alter the integrated Wigner
negativity. Changes in the interference structure and negative regions arise
from the environmental dynamics, with~dissipative transport contained in
$M(t)$ and Gaussian broadening determined by the bath-induced covariance
matrix $\Sigma(t)$. This separation provides a useful framework for
interpreting future CNT tomography experiments in which coherent actuation
and environmentally induced loss of nonclassical structure occur
simultaneously.

\section{Wigner Negativity Under Driven Finite-Memory~Dynamics}
\label{sec:negativity}

The integrated Wigner negativity $\mathcal N_W(t)$ introduced in
Section~\ref{sec:cat} provides a phase space measure of Wigner
nonclassicality that complements the fringe attenuation coefficient
$a(t)$ \cite{KenfackZyczkowski2004}. Here, we establish two properties
that follow directly from the Gaussian propagator: invariance under a
deterministic AFM displacement and an upper bound under the
reservoir-induced Gaussian evolution. {This section does not introduce an additional source of quantum noise. Rather, it uses the classical and full quantum covariance matrices derived in the preceding sections to determine how reservoir fluctuations affect nonclassical structure in the Wigner function. In~particular, replacing the classical covariance by the full quantum covariance within the same propagator isolates the effects of zero-point and low-temperature quantum fluctuations on the integrated negative part of the Wigner~distribution.}

Let $W_F(\mathbf z,t)$ denote the Wigner function evolved with a prescribed
force $F(t)$, and~let $W_0(\mathbf z,t)$ denote the evolution from the same
initial state and with the same reservoir, but~with $F(t)=0$. As~shown for
linearly driven QBM, a~prescribed classical force changes the phase space
displacement while leaving the bath-induced covariance unchanged~\cite{FlemingRouraHu2011,QiuQuan2021}. Therefore,
\begin{equation}
\begin{aligned}
P_F(\mathbf z,\mathbf z';t)
&=P_0(\mathbf z-\boldsymbol{\eta}(t),\mathbf z';t),\\
W_F(\mathbf z,t)
&=W_0(\mathbf z-\boldsymbol{\eta}(t),t).
\end{aligned}
\label{eq:Wshift}
\end{equation}
Since a translation in phase space has unit Jacobian,
\begin{equation}
\mathcal N_W^{(F)}(t)
=
\mathcal N_W^{(0)}(t).
\label{eq:NegDriveInvariant}
\end{equation}
Thus, a~prescribed noiseless AFM force changes the location of the Wigner
distribution in phase space but does not change its integrated negative
weight. Any change in the Wigner negativity therefore arises from the
reservoir-induced evolution rather than from the deterministic displacement
itself.

A second property follows from the positivity and normalization of the
Gaussian transition kernel in the exact linear-QBM propagator~\cite{FordOConnell2001,FlemingRouraHu2011}. For~$t>0$, whenever the
covariance matrix entering the propagator is nonsingular,
\begin{align}
\int\dd^2z\,|W(\mathbf z,t)|
&=
\int\dd^2z\,
\left|
\int\dd^2z'\,
P(\mathbf z,\mathbf z';t)W(\mathbf z',0)
\right|
\nonumber\\
&\leq
\int\dd^2z'\,
|W(\mathbf z',0)|
\int\dd^2z\,P(\mathbf z,\mathbf z';t)
\nonumber\\
&=
\int\dd^2z'\,|W(\mathbf z',0)|.
\label{eq:WignerL1Bound}
\end{align}
Using the definition of $\mathcal N_W(t)$ in Equation~\eqref{eq:Wneg}, it
follows that
\begin{equation}
\mathcal N_W(t)
\leq
\mathcal N_W(0).
\label{eq:NegBound}
\end{equation}
The singular $t\rightarrow0$ limit is understood continuously, with~the
Gaussian transition kernel approaching the corresponding phase space
delta~distribution.

Equation~\eqref{eq:NegBound} compares the Wigner negativity at time $t$
with that of the initial state. It should not be interpreted as a proof
that $\mathcal N_W(t)$ decreases monotonically between arbitrary
intermediate times. For~non-Markovian dynamics, the~map connecting two
intermediate times need not be representable by a positive, normalized
Gaussian transition kernel, even though the map from the specified initial
time to $t$ has the form used above. Consequently, decreases followed by
partial revivals of $\mathcal N_W(t)$ at intermediate times are not
excluded by Equation~\eqref{eq:NegBound}.

For the two-packet state considered in Section~\ref{sec:cat},
$\mathcal N_W(t)$ is obtained directly from the propagated Wigner function
and requires no additional dynamical equation. The~attenuation coefficient
$a(t)$ characterizes the suppression of interference in the position
marginal, whereas $\mathcal N_W(t)$ quantifies the negative weight of the
complete phase space distribution. Comparing the predictions obtained with
the classical-noise covariance of Section~\ref{sec:classicalcov} and the full
quantum covariance of Section~\ref{sec:quantumdrude} therefore provides a
useful diagnostic of how zero-point and low-temperature quantum
fluctuations modify the persistence of nonclassical phase space structure.
Where the classical-noise approximation does not satisfy the criteria discussed in Sections~\ref{sec:validity} and~\ref{sec:physicality}, its negativity prediction provides a comparative benchmark for identifying
and quantifying quantum fluctuation effects, rather than a quantitative physical
prediction.

\section{Discussion}
\label{sec:discussion}

The results presented here establish an explicit analytical framework for
the finite-memory Gaussian phase space dynamics of a driven bound oscillator
within linear QBM. Exact formulations by Hu, Paz, and~Zhang; Ford and O'Connell; and~later Fleming, Roura, and~Hu established that a harmonic system coupled linearly
to a Gaussian environment admits an exact reduced description in terms of
dissipative response functions and environmental fluctuations~\cite{HPZ1992,FordOConnell2001,FlemingRouraHu2011}. The~present analysis
retains this structure while focusing on a bound mechanical mode with an
explicitly resolved reservoir memory time. This distinction is important for
nanomechanical resonators, where oscillatory dynamics and reservoir relaxation
occur simultaneously and can be comparable on experimentally relevant timescales.

From a mathematical perspective, the~main result is an explicit reduction of
the finite-memory bound-oscillator problem to a pole-based representation.
For the Drude kernel, the~retarded Green function is determined by the three
roots of a characteristic cubic, and~the finite-time classical covariance
matrix is reduced from double-time integrals to finite sums of elementary
functions of these poles. Retaining the full quantum fluctuation spectrum
leads to the same response-pole structure combined with a Matsubara frequency
representation of the thermal correlations. For~a finite sum of $N$
exponential memory terms, the~construction generalizes to a rational
Laplace-domain response governed by a polynomial of degree $N+2$, while the
classical covariance remains an explicit finite sum. The~same framework also
yields the uncertainty-based consistency condition
$k_BT\geq\hbar\omega_0/2$ for the asymptotic classical covariance and permits
direct proofs of both the invariance of integrated Wigner negativity under
deterministic phase space translations and its upper bound under the
positive Gaussian propagation from the initial time. These results provide
an analytically tractable formulation of finite-memory QBM that connects
nonlocal response, covariance evolution, and~phase space nonclassicality
within a common mathematical structure~\cite{FordOConnell2001,FlemingRouraHu2011,GrabertSchrammIngold}.

An important physical consequence of finite memory is that dissipation and
fluctuations cannot be interpreted solely through instantaneous damping
rates. The~reservoir correlation time enters both the retarded response and
the fluctuation covariance, so changes in memory affect both the mechanical
motion and the spreading of the reduced state~\cite{GrabertSchrammIngold,FlemingRouraHu2011}. In~the Drude model, this
relation is particularly transparent because the response and noise
correlations admit compatible exponential representations. The~multi-exponential extension shows that the same mathematical organization
can accommodate more than one environmental relaxation time. Therefore, the~finite sums of
exponentials provide a useful class of non-Markovian models in which several
memory scales can be incorporated without losing analytical tractability~\cite{Park2024Pseudomode,XuStockburgerAnkerhold2026}.

The comparison between classical and quantum fluctuations is especially
relevant in the low-temperature CNT regime. Classical thermal noise can
provide a useful approximation when the response-weighted environmental
frequencies are sufficiently small compared with the thermal scale, but~satisfying the uncertainty relation is a weaker requirement than accurately
reproducing the full quantum fluctuation spectrum~\cite{FordLewisOConnell1988,GrabertSchrammIngold}. The~consistency criterion
derived here should therefore be interpreted as a test of whether the
asymptotic classical covariance can represent a physical one-mode quantum
state at all, rather than as a general criterion for the quantitative
validity of a classical-noise bath model. At~sufficiently high mechanical
frequency or sufficiently low-temperature, zero-point and finite-temperature
quantum corrections must be retained. This distinction is consistent with
recent analyses emphasizing the importance of quantum colored noise and
high-frequency quantum fluctuations in low-temperature and finite-cutoff QBM~\cite{KondaurovPolyakov2026,Pleasance2026}.

The separation between deterministic forcing and environmental evolution also
has a direct experimental meaning. For~linear QBM with a prescribed classical
force, the~drive changes the first moments through the deterministic phase
space displacement, whereas the bath-induced covariance is independent of
the driving waveform~\cite{FlemingRouraHu2011,QiuQuan2021}. In~the present harmonic CNT model, the~AFM force therefore controls the displacement of the state, while the
reservoir determines dissipative transport, fluctuation-induced broadening,
and the suppression of interference structure. This distinction is useful
for tomography because motion of the reconstructed Wigner distribution and
loss of its nonclassical features need not originate from the same physical
mechanism. The~integrated Wigner negativity provides a global phase space
measure of the negative weight of the quasiprobability distribution~\cite{KenfackZyczkowski2004}, and~its invariance under a prescribed noiseless
displacement makes this separation particularly transparent. The~two-packet
attenuation factor provides a complementary measure associated with the
position marginal, while the no-drive limit isolates the reservoir-induced
evolution and recovers the standard Gaussian QBM propagation considered by
Ford and O'Connell~\cite{FordOConnell2001}.

For the proposed CNT platform, the~present calculation provides an analytical
harmonic baseline for the more general control problem. The~AFM-based
architecture proposed in Ref.~\cite{Staii2026} exploits Duffing/Kerr
anharmonicity to obtain spectrally selective manipulation of the lowest
vibrational states. Once such nonlinear terms become dynamically important,
the Wigner evolution is no longer determined entirely by first and second
moments, and~the simple Gaussian propagation developed here no longer closes
the dynamics. The~harmonic solution therefore provides a well-defined
baseline against which nonlinear corrections to the response, phase space
evolution, and~decoherence can be~identified.

{Nonlinear mechanical response is a generic feature of CNT resonators and
has also been observed in recent experiments in which a Duffing description
is required}
\cite{Fedele2025}{. For~weak Duffing or Kerr nonlinearities, a~natural
extension of the present calculation would be to use the exact finite-memory
Gaussian solution as the unperturbed dynamics and treat the nonlinear
potential perturbatively in the Wigner evolution. This approach would
retain the non-Markovian reservoir response developed here while generating
systematic corrections to the harmonic phase space dynamics. When the
nonlinearity is too strong for such an expansion, nonequilibrium
functional-integral methods, including the Schwinger--Keldysh formulation,
or numerical stochastic phase space methods provide more general approaches
to nonlinear dissipative dynamics}
\cite{Kamenev2023}{. Extending the present finite-memory treatment along
these lines would allow for the Duffing/Kerr regime relevant to coherent CNT
control to be addressed beyond the Gaussian approximation.}

The environmental model introduces a second limitation. The~Drude cutoff
$\omega_c$, and~more generally the parameters $\{\gamma_a,\kappa_a\}$ of the
multi-exponential memory kernel, are not presently fixed by the available CNT
mechanical parameters. Quantitative comparison with experiment will therefore
require either microscopic modeling of the dominant damping mechanisms or
independent characterization of the relevant reservoir spectrum. The~explicit pole representation is useful in this respect because the reservoir
relaxation scales enter the response and covariance matrix directly and can, in~principle, be constrained by measured mechanical response and fluctuation
data. {Experimentally, these memory parameters could be constrained by combining
frequency-domain measurements of the complex mechanical susceptibility with
time-domain ring down measurements} \cite{Schneider2014}{. A~simple exponential ring down primarily determines
the near-resonant damping rate, whereas resolving transient
behavior or measuring the response over a sufficiently broad frequency range
would be required to determine a finite cutoff}
$\omega_c$
{or the relaxation rates}
$\gamma_a${. Measurements of the equilibrium displacement-noise spectrum
would provide an additional constraint through the fluctuation--dissipation
relation. Microscopic models of support and clamping losses, coupling to
substrate phonons through the CNT supports, and~other device-specific
dissipation mechanisms could then be used to motivate the functional form and
characteristic scales of the memory kernel.}

The assumed initial preparation also represents an important modeling
consideration. The~factorized oscillator--bath state provides continuity
with the HPZ and Ford--O'Connell formulations and leads directly to the
finite-time propagator used here~\cite{HPZ1992,FordOConnell2001,FlemingRouraHu2011}. A~CNT oscillator that
has equilibrated with its environment before the control sequence begins
may retain initial system--environment correlations. Correlated initial
preparations can modify the short time dynamics and are particularly
relevant when low-temperature and strong system--bath correlation effects
are important~\cite{RomeroPaz1997,KondaurovPolyakov2026}. Extending the
present bound-oscillator treatment to correlated initial conditions would
therefore be a natural direction toward a more complete experimental
description.

Taken together, these results provide an analytically explicit framework for
determining how finite reservoir memory modifies the response, fluctuations,
and phase space coherence of a driven bound oscillator. For~CNT tomography,
the framework separates coherent AFM actuation from reservoir-induced
dissipation and fluctuation effects, quantifies when a classical-noise
description becomes inadequate, and~provides complementary measures of
coherence loss through interference attenuation and integrated Wigner
negativity. More broadly, the~pole-based construction offers a tractable
route for studying driven Gaussian open systems with structured memory while
retaining direct connections between mathematical response theory,
finite-time covariance dynamics, and~experimentally observable phase space
behavior.

\section{Conclusions}
\label{sec:conclusions}

We have developed an explicit finite-time phase space description of a
driven, bound quantum Brownian oscillator with finite environmental memory
and applied it to the harmonic flexural mode of an AFM-driven suspended
carbon nanotube. For~a Drude environment, the~response is expressed through
the characteristic poles of the bound oscillator, yielding explicit
finite-sum formulas for the classical covariance matrix and a representation of the full quantum covariance in terms of response poles and Matsubara frequencies that retains zero-point and low-temperature fluctuations. The~construction
extends to reservoirs described by finite sums of exponential memory
kernels. The~framework also provides an uncertainty-based consistency test
for the asymptotic classical covariance and phase space diagnostics that
separate deterministic displacement from reservoir-induced changes in
nonclassical~structure.

From a mathematical perspective, the~pole representation reduces the nonlocal response problem and the associated double-time covariance integrals to explicit algebraic and exponential forms. The~extension to multiple exponential memory components preserves this analytic structure while allowing for several reservoir relaxation times to be treated systematically. These results provide a tractable framework linking finite-memory response, covariance evolution, and~Wigner dynamics in driven Gaussian open~systems.

Beyond the specific carbon nanotube applications, the~results provide an analytical
framework for studying quantum Brownian motion with structured environments
in regimes where environmental memory, low-temperature fluctuations, and~decoherence must be treated simultaneously. By~making explicit the
connection between reservoir correlation times, dissipative response,
fluctuation-induced spreading, and~the suppression of Wigner function
interference, the~approach helps clarify how non-Markovian environments
influence coherence and the emergence of classical behavior in bound
oscillators. The~extension to multiple relaxation times further broadens
its relevance to structured reservoirs encountered in nanomechanical and
mesoscopic~systems.

For nanoresonators, the~framework provides a direct route from mechanical
and environmental parameters to finite-time quantum dynamics and coherence
loss. In~particular, it provides a quantitative basis for separating
coherent control from environmental decoherence in Wigner function
tomography and related state-reconstruction protocols. The~harmonic
solution developed here can serve as an analytical benchmark for more
complex open system models and as a starting point for future studies
incorporating nonlinear mechanical dynamics, correlated initial states,
and experimentally characterized reservoir spectra. The~present results
therefore provide a theoretical basis for interpreting future experiments
on quantum state preparation and tomography in quantum~nanoresonators.

\section*{Funding}
This
 research was funded by a Tufts University Faculty Research Award (C.S. (Cristian Staii)) and Tufts Summer Scholars Program (Conrad Silio).

\section*{Data Availability}
{The}  original contributions presented in this study are included in the
article. Further inquiries can be directed to the corresponding author.

\section*{Conflicts of Interest}
The authors declare no conflicts of interest. The~funders had no role in the design
of the study; in the collection, analyses, or~interpretation of data; in the writing of the manuscript; or
in the decision to publish the~results.


\clearpage
\onecolumngrid

\begin{center}
{\bfseries\large Supplemental Material for\\[3pt]
Quantum Fluctuations and Wigner Dynamics of a Carbon Nanotube Oscillator Driven by an Atomic Force Microscope}
\end{center}

\setcounter{section}{0}
\setcounter{subsection}{0}
\setcounter{equation}{0}
\setcounter{table}{0}
\renewcommand{\thesection}{\Alph{section}}
\makeatletter
\@addtoreset{equation}{section}
\@addtoreset{table}{section}
\makeatother
\renewcommand{\theequation}{\Alph{section}\arabic{equation}}
\renewcommand{\thetable}{\Alph{section}\arabic{table}}

\section{Derivation of the Classical Drude Covariance~Matrix}
\label{app:covariance}

Starting from the general covariance integrals in
Equations~\eqref{eq:sigmaqqgeneral}--\eqref{eq:sigmappgeneral}, make the
substitutions
\begin{equation}
u=t-s,
\qquad
v=t-s'.
\end{equation}
Since $|s-s'|=|u-v|$, the~classical Drude force correlation function in
Equation~\eqref{eq:classicalFDT} contributes the factor
$K\ee^{-\omega_c|u-v|}$. Using the pole representation of the Green
function in Equation~\eqref{eq:Gpoles}, the~double integrals therefore contain
terms proportional to
\begin{equation}
\ee^{\lambda_i u+\lambda_j v-\omega_c|u-v|},
\qquad
i,j=1,2,3,
\end{equation}
where $\lambda_i$ are the characteristic roots introduced in
Equation~\eqref{eq:Gpoles}.

To evaluate $I_{ij}(t)$ defined in Equation~\eqref{eq:Iij}, split the square
$0\leq u,v\leq t$ into the two triangular regions $u\geq v$ and
$v\geq u$. The~common boundary $u=v$ has zero measure and therefore does
not contribute separately. In~the region $u\geq v$,
\begin{align}
I_{ij}^{(u\geq v)}(t)
&=
\int_0^t\dd u\,
\ee^{(\lambda_i-\omega_c)u}
\int_0^u\dd v\,
\ee^{(\lambda_j+\omega_c)v}
\nonumber\\
&=
\frac{
\mathcal E(\lambda_i+\lambda_j,t)
-\mathcal E(\lambda_i-\omega_c,t)
}{
\lambda_j+\omega_c
},
\label{eq:IijTriangle}
\end{align}
where $\mathcal E(a,t)$ is defined in Equation~\eqref{eq:Edef}. The~contribution
from the region $v\geq u$ is obtained by interchanging $i$ and $j$,
\begin{equation}
I_{ij}^{(v\geq u)}(t)
=
\frac{
\mathcal E(\lambda_i+\lambda_j,t)
-\mathcal E(\lambda_j-\omega_c,t)
}{
\lambda_i+\omega_c
}.
\end{equation}
Adding the two triangular contributions gives Equation~\eqref{eq:Iij}.

Using the pole expansion in Equation~\eqref{eq:Gpoles},
\begin{align}\label{eq:covariancePoleProducts}
G(u)G(v)
&=
\sum_{i,j=1}^{3}
c_i c_j
\ee^{\lambda_i u+\lambda_j v},
\nonumber\\
\frac{m}{2}
\left[
G(u)\dot G(v)+\dot G(u)G(v)
\right]
&=
\frac{m}{2}
\sum_{i,j=1}^{3}
c_i c_j(\lambda_i+\lambda_j)
\ee^{\lambda_i u+\lambda_j v},
\\
m^2\dot G(u)\dot G(v)
&=
m^2
\sum_{i,j=1}^{3}
c_i c_j\lambda_i\lambda_j
\ee^{\lambda_i u+\lambda_j v}.
\nonumber
\end{align}
Multiplication by the classical Drude correlator
$K\ee^{-\omega_c|u-v|}$, with~$K$ defined in
Equation~\eqref{eq:classicalFDT}, and~integration over $u$ and $v$ therefore
produce the common factor $I_{ij}(t)$ in each covariance element. This
directly yields Equations~\eqref{eq:SigmaqqClosed}--\eqref{eq:SigmappClosed}.

\section{Derivation of the Two-Packet Attenuation~Factor}
\label{app:cat}

For the derivation of the position marginal, it is convenient to introduce
scaled Fourier variables $(Q,P)$ through
\begin{equation}
\widetilde W(Q,P;t)
=
\int\dd q\,\dd p\,
W(q,p;t)
\exp\!\left[
-\frac{\ii}{\hbar}(Pq+Qp)
\right].
\label{eq:dimensionalFourier}
\end{equation}
Here, $Q$ has dimensions of length and is conjugate to momentum, while $P$
has dimensions of momentum and is conjugate to position. For~the initial
two-packet state in Equation~\eqref{eq:catpsi}, direct transformation of
Equation~\eqref{eq:catW0} gives
\begin{equation}
\widetilde W(Q,P;0)
=
\frac{1}{1+S}
\exp\!\left[
-\frac{Q^2}{8\sigma^2}
-\frac{\sigma^2P^2}{2\hbar^2}
\right]
\left[
\cos\!\left(\frac{Pd}{2\hbar}\right)
+
S\cosh\!\left(\frac{Qd}{4\sigma^2}\right)
\right].
\label{eq:catFourier}
\end{equation}

The position marginal
$\mathcal P(q,t)=\int\dd p\,W(q,p;t)$
is obtained by setting the Fourier variable $Q$, which is conjugate to
momentum, equal to zero after propagation. Using the phase space solution
of Section~\ref{sec:solution} then gives
\begin{equation}
\mathcal P(q,t)
=
\frac{1}{2\pi\hbar}
\int_{-\infty}^{\infty}\dd s\,
\widetilde W\!\left(G(t)s,m\dot G(t)s;0\right)
\exp\!\left[
-\frac{\Sigma_{qq}(t)s^2}{2\hbar^2}
+\frac{\ii s[q-\eta_q(t)]}{\hbar}
\right],
\label{eq:catmarginal}
\end{equation}
where the integration variable $s$ has dimensions of momentum. The~Gaussian
width appearing in this integral is the propagated packet variance
$\langle\Delta q^2(t)\rangle$ given in
Equation~\eqref{eq:packetvariance}.

The cosine term in Equation~\eqref{eq:catFourier} generates the two Gaussian
packet contributions, whose centers are
$\eta_q(t)\pm m\dot G(t)d/2$. The~hyperbolic-cosine term generates the
interference contribution. Following the two-packet visibility measure used
by Ford and O'Connell~\cite{FordOConnell2001}, the~attenuation factor is
defined as the ratio of the interference coefficient to twice the geometric
mean of the two packet coefficients. Evaluation of the Gaussian integral
gives
\begin{equation}
a(t)
=
S
\exp\!\left[
\frac{\hbar^2G^2(t)d^2}
{32\sigma^4\langle\Delta q^2(t)\rangle}
+
\frac{m^2\dot G^2(t)d^2}
{8\langle\Delta q^2(t)\rangle}
\right].
\label{eq:attenuationIntermediate}
\end{equation}
Using
$S=\exp[-d^2/(8\sigma^2)]$
together with Equation~\eqref{eq:packetvariance}, the~exponent simplifies
directly to
\begin{equation}
-\frac{d^2\Sigma_{qq}(t)}
{8\sigma^2\langle\Delta q^2(t)\rangle},
\end{equation}
thereby recovering Equation~\eqref{eq:attenuation}. The~common deterministic
displacement $\eta_q(t)$ cancels from this ratio, consistent with
the fact that a prescribed force translates both packets by the same~amount.

\section{Numerical and Limiting-Case~Validation}
\label{app:numericalchecks}

We perform two independent checks of the analytical results. First, we
compare the classical Drude covariance derived in
Section~\ref{sec:classicalcov} with an equivalent time-domain calculation
that does not use the response poles $\lambda_i$ or the integrals
$I_{ij}(t)$. This also provides an independent check of the pole
representation of the retarded Green function derived in
Section~\ref{sec:drude}. Second, we take the free-particle and large-cutoff
limits and verify convergence to the high-temperature Ohmic result used
in the Ford--O'Connell two-packet analysis~\cite{FordOConnell2001}.

For the exponential Drude memory kernel in Equation~\eqref{eq:drudemu}, define
the auxiliary variable
\begin{equation}
Y(t)
=
\int_0^t\dd s\,
\ee^{-\omega_c(t-s)}
\dot X(s).
\label{eq:Yembedding}
\end{equation}
The bath-driven displacement $X(t)$ defined in
Equation~\eqref{eq:XandEta} then satisfies
\begin{equation}
\dot Y
=
\dot X-\omega_cY,
\qquad
\ddot X
=
-\omega_0^2X
-2\Gamma_0\omega_cY
+\frac{F_B}{m}.
\label{eq:Xembedding}
\end{equation}
Thus, the~exponential memory integral is represented by the additional
dynamical variable $Y(t)$, providing a time-local embedding of the
classical finite-memory~problem.

The exponentially correlated classical bath force associated with
Equation~\eqref{eq:classicalFDT} can be generated by the
Ornstein--Uhlenbeck process
\begin{equation}
\dd F_B
=
-\omega_cF_B\,\dd t
+
\sqrt{2\omega_cK}\,\dd W_t,
\label{eq:OUforce}
\end{equation}
where $W_t$ is a standard Wiener process and
$K=2m\Gamma_0\omega_c k_BT$ is the coefficient defined in
Equation~\eqref{eq:classicalFDT}. Choosing $F_B(0)$ from its stationary
Gaussian distribution with variance $K$~gives
\begin{equation}
\langle F_B(t)F_B(t')\rangle
=
K\ee^{-\omega_c|t-t'|},
\end{equation}
which is precisely the classical Drude force correlation function in
Equation~\eqref{eq:classicalFDT}.

Introduce the auxiliary state vector
\begin{equation}
\mathbf r
=
\begin{pmatrix}
X &
\dot X &
Y &
F_B
\end{pmatrix}^{\mathsf T}.
\end{equation}
Equations~\eqref{eq:Xembedding} and~\eqref{eq:OUforce} can then be written
as the linear stochastic system
\begin{equation}
\dd\mathbf r
=
\mathcal A\mathbf r\,\dd t
+
\mathcal B\,\dd W_t,
\end{equation}
where
\begin{equation}
\mathcal A
=
\begin{pmatrix}
0 & 1 & 0 & 0 \\
-\omega_0^2 & 0 & -2\Gamma_0\omega_c & 1/m \\
0 & 1 & -\omega_c & 0 \\
0 & 0 & 0 & -\omega_c
\end{pmatrix},
\qquad
\mathcal B
=
\begin{pmatrix}
0 \\
0 \\
0 \\
\sqrt{2\omega_cK}
\end{pmatrix}.
\label{eq:embeddingMatrices}
\end{equation}

All four variables have zero mean. Their covariance matrix
\begin{equation}
\mathcal C(t)
=
\left\langle
\mathbf r(t)\mathbf r^{\mathsf T}(t)
\right\rangle
\end{equation}
therefore satisfies the linear covariance equation
\begin{equation}
\dot{\mathcal C}
=
\mathcal A\mathcal C
+
\mathcal C\mathcal A^{\mathsf T}
+
\mathcal B\mathcal B^{\mathsf T},
\qquad
\mathcal C(0)
=
\operatorname{diag}(0,0,0,K).
\label{eq:embeddingLyapunov}
\end{equation}
The bath-induced covariance elements appearing in
Section~\ref{sec:classicalcov} follow directly as
\begin{equation}
\Sigma_{qq}(t)=\mathcal C_{11}(t),
\qquad
\Sigma_{qp}(t)=m\mathcal C_{12}(t),
\qquad
\Sigma_{pp}(t)=m^2\mathcal C_{22}(t).
\label{eq:embeddingCovariances}
\end{equation}
This calculation is independent of the characteristic roots $\lambda_i$
and the finite-sum expressions involving $I_{ij}(t)$.

For the parameter choice
$\Gamma_0=\omega_0/(2Q)$ with $Q=50$ and
$\omega_c/\omega_0=10$, all three covariance elements obtained from
Equation~\eqref{eq:embeddingLyapunov} agree with the finite pole-sum results
in Equations~\eqref{eq:SigmaqqClosed}--\eqref{eq:SigmappClosed} to within
$2\times10^{-10}$ over
\[
0\leq\omega_0t\leq8\pi,
\]
when each difference is normalized by the maximum magnitude of the
corresponding analytical quantity over this interval. A~direct numerical
solution of the Green-function Equation~\eqref{eq:Gdef} agrees with the
pole representation in Equation~\eqref{eq:Gpoles} to within
$2\times10^{-12}$ using the same normalization. As~an additional check,
direct numerical evaluation of the original double-time integrals in
Equations~\eqref{eq:sigmaqqgeneral}--\eqref{eq:sigmappgeneral} at selected
times gives relative differences below $3\times10^{-7}$ from the
finite-sum expressions. These comparisons independently validate both
the response-pole construction of Section~\ref{sec:drude} and the closed
classical covariance formulas of Section~\ref{sec:classicalcov}, including
the derivation summarized in Appendix~\ref{app:covariance}.

For the second check, set $\omega_0=0$, so that the bound oscillator
reduces to a free Brownian particle, and~consider the Ohmic limit
$\omega_c\rightarrow\infty$. Defining
\begin{equation}
\zeta=2m\Gamma_0,
\label{eq:zetaFO}
\end{equation}
the corresponding retarded Green function is
~\cite{FordOConnell2001}
\begin{equation}
G_{\mathrm{FO}}(t)
=
\frac{1-\ee^{-\zeta t/m}}{\zeta}.
\label{eq:GFO}
\end{equation}

{Its time derivative, which enters the propagated packet variance in
Equation~}\eqref{eq:packetvariance}{, is}
\begin{equation}
\dot G_{\mathrm{FO}}(t)
=
\frac{1}{m}\ee^{-\zeta t/m}.
\label{eq:GFOdot}
\end{equation}
In the classical high-temperature noise limit, the~bath-induced position
variance is
~\cite{FordOConnell2001}
\begin{equation}
\Sigma_{qq}^{\mathrm{FO}}(t)
=
\frac{k_BT}{\zeta}
\left[
2t
-
\frac{2m}{\zeta}
\left(
1-\ee^{-\zeta t/m}
\right)
-
\frac{m}{\zeta}
\left(
1-\ee^{-\zeta t/m}
\right)^2
\right].
\label{eq:SigmaFO}
\end{equation}
{Together, Equations~}\eqref{eq:GFO}--\eqref{eq:SigmaFO}{, when substituted into Equations~}\eqref{eq:packetvariance}{ and~}\eqref{eq:attenuation}{, reproduce the high-temperature factorized-state result used in the Ford--O'Connell two-packet analysis}~\cite{FordOConnell2001}{.} This limiting case therefore
also tests the attenuation construction developed in
Section~\ref{sec:cat} and Appendix~\ref{app:cat}.

We compare the Ohmic expressions with the finite-cutoff classical Drude
calculation. Since $\omega_0=0$ in the free-particle limit, the~cutoff is
expressed relative to the damping scale $\Gamma_0$. Table~\ref{tab:FO_validation}
lists the largest differences over
\begin{equation}
  0\leq\frac{\zeta t}{m}\leq4.
\end{equation}

The differences in $G$ and $\Sigma_{qq}$ are normalized by the maximum
magnitude of the corresponding Ford--O'Connell quantity over this
interval, while the attenuation column gives the maximum absolute
difference.

\begin{table}[t]
\label{tab:FO_validation}
\begin{tabular}{cccc}
\toprule
\boldmath{$\omega_c/\Gamma_0$}
&
\boldmath{$\displaystyle
\frac{\max|\Delta G|}
{\max|G_{\mathrm{FO}}|}
$}
&
\boldmath{$\displaystyle
\frac{\max|\Delta\Sigma_{qq}|}
{\max|\Sigma_{qq}^{\mathrm{FO}}|}
$}
&
\boldmath{$\displaystyle
\max|\Delta a|
$}
\\
\midrule
$200$  & $3.77\times10^{-3}$ & $3.30\times10^{-3}$ & $4.20\times10^{-2}$ \\
$1000$ & $7.50\times10^{-4}$ & $6.60\times10^{-4}$ & $8.88\times10^{-3}$ \\
$5000$ & $1.50\times10^{-4}$ & $1.32\times10^{-4}$ & $1.77\times10^{-3}$ \\
\bottomrule
\end{tabular}
\caption{Convergence of the classical-noise free-particle Drude calculation
to the Ford--O'Connell Ohmic result. The~calculation uses the illustrative
units $m=k_BT=\hbar=1$, together with $\Gamma_0=0.01$, $d=3.5$, and~$\sigma=0.6$. These parameters are used here to test the mathematical
limiting behavior of the classical-noise formulas rather than to represent
a physical high-temperature CNT parameter set.}
\end{table}

The
 differences decrease systematically as the cutoff is increased.
Increasing $\omega_c/\Gamma_0$ from $200$ to $1000$, and~again from
$1000$ to $5000$, reduces the errors in both $G$ and $\Sigma_{qq}$ by
approximately a factor of five. The~attenuation difference follows the
same trend, confirming convergence of the finite-cutoff Drude calculation
toward the free-particle Ohmic~limit.

\end{document}